\documentclass[journal]{IEEEtran}
\usepackage{verbatim}
\usepackage{balance}
\usepackage{multirow}
\usepackage{booktabs}
\usepackage{amsfonts}
\usepackage{amsmath}
\usepackage{amsmath,cases}
\usepackage{amssymb}
\usepackage{graphicx}
\usepackage{cite}
\usepackage{epsfig}
\usepackage{url}
\usepackage{algorithm}
\usepackage{algorithmicx, algpseudocode}
\usepackage{epstopdf}
\usepackage{color}
\usepackage[tight]{subfigure}
\usepackage{mathrsfs}
\usepackage[justification=centering]{caption}
\usepackage{float}
\usepackage{diagbox}

\newcommand{\ds}{\displaystyle}

\newcommand{\la}{\langle}
\newcommand{\ra}{\rangle}

\newcommand{\clC}{{\cal C}}

\newcommand{\clK}{{\cal K}}

\newcommand{\bw}{\mathbf{w}}
\newcommand{\bv}{\mathbf{v}}

\newcommand{\clH}{{\cal H}}

\newcommand{\bW}{\mathbf{W}}

\newcommand{\ak}{a^{(\kappa)}}
\newcommand{\Bk}{B^{(\kappa)}}
\newcommand{\Ck}{C^{(\kappa)}}

\newcommand{\gammak}{\gamma^{(\kappa)}}

\newcommand{\clQ}{\mathcal{Q}}
\newcommand{\Wko}{W^{(\kappa+1)}}

\newcommand{\Wk}{W^{(\kappa)}}

\newcommand{\ck}{c^{(\kappa)}}

\newcommand{\wk}{w^{(\kappa)}}
\newcommand{\wko}{w^{(\kappa+1)}}
\newcommand{\wka}{w^{a,(\kappa)}}
\newcommand{\wkoa}{w^{a,(\kappa+1)}}
\newcommand{\wke}{w^{e,(\kappa)}}
\newcommand{\wkoe}{w^{e,(\kappa+1)}}
\newcommand{\varphik}{\varphi^{(\kappa)}}
\newcommand{\tvarphik}{\tilde{\varphi}^{(\kappa)}}
\newcommand{\hke}{h^{e,(\kappa)}}
\newcommand{\hka}{h^{a,(\kappa)}}
\newcommand{\aka}{a^{a,(\kappa)}}
\newcommand{\bka}{b^{a,(\kappa)}}
\newcommand{\cka}{c^{a,(\kappa)}}

\newcommand{\ake}{a^{e,(\kappa)}}
\newcommand{\bke}{b^{e,(\kappa)}}
\newcommand{\cke}{c^{e,(\kappa)}}

\newcommand{\Cka}{C^{a,(\kappa)}}
\newcommand{\Cke}{C^{e,(\kappa)}}

\newcommand{\Qek}{\clQ^{e,(\kappa)}}
\newcommand{\Qak}{\clQ^{a,(\kappa)}}
\newcommand{\tQek}{\tilde{\clQ}^{e,(\kappa)}}
\newcommand{\tQak}{\tilde{\clQ}^{a,(\kappa)}}
\newcommand{\hQek}{\hat{\clQ}^{e,(\kappa)}}
\newcommand{\hQak}{\hat{\clQ}^{a,(\kappa)}}
\newcommand{\rka}{r^{a,(\kappa)}}
\newcommand{\rke}{r^{e,(\kappa)}}
\newcommand{\tbw}{\tilde{\bw}}
\newcommand{\hbw}{\hat{\bw}}
\newcommand{\hr}{\hat{r}}
\newcommand{\hvarphi}{\hat{\varphi}}
\newcommand{\hwk}{\hat{w}^{(\kappa)}}
\newcommand{\hwko}{\hat{w}^{(\kappa+1)}}
\newcommand{\hwka}{\hat{w}^{a,(\kappa)}}
\newcommand{\hwke}{\hat{w}^{e,(\kappa)}}
\newcommand{\hwkoa}{\hat{w}^{a,(\kappa+1)}}
\newcommand{\hwkoe}{\hat{w}^{e,(\kappa+1)}}
\newcommand{\hgammak}{\hat{\gamma}^{(\kappa)}}
\newcommand{\hvarphik}{\hat{\varphi}^{(\kappa)}}

\newcommand{\twke}{\tilde{w}^{e,(\kappa)}}

\newcommand{\hhek}{\hat{h}^{e,(\kappa)}}
\newcommand{\hhak}{\hat{h}^{a,(\kappa)}}
\newcommand{\bva}{\mathbf{v}^{a}}
\newcommand{\bve}{\mathbf{v}^{e}}

\newcommand{\tih}{\tilde{h}}
\newcommand{\hh}{\hat{h}}
\newcommand{\hvarphiak}{\hat{\varphi}^{a,(\kappa)}}
\newcommand{\hvarphiek}{\hat{\varphi}^{e,(\kappa)}}
\newcommand{\vka}{v^{a,(\kappa)}}
\newcommand{\vkoa}{v^{a,(\kappa+1)}}
\newcommand{\vke}{v^{e,(\kappa)}}
\newcommand{\vkoe}{v^{e,(\kappa+1)}}
\newcommand{\clHek}{\clH^{e,(\kappa)}}
\newcommand{\clHak}{\clH^{a,(\kappa)}}
\newcommand{\hrak}{\hat{r}^{a,(\kappa)}}
\newcommand{\hrek}{\hat{r}^{e,(\kappa)}}
\newcommand{\Bka}{B^{a,(\kappa)}}
\newcommand{\Bke}{B^{e,(\kappa)}}
\newcommand{\hbka}{\hat{b}^{a,(\kappa)}}
\newcommand{\hbke}{\hat{b}^{e,(\kappa)}}
\newcommand{\twkoa}{\tilde{w}^{a,(\kappa+1)}}

\newcommand{\hrhoak}{\hat{\rho}^{a,(\kappa)}}
\newcommand{\hrhoek}{\hat{\rho}^{e,(\kappa)}}
\newcommand{\Col}{{\sf Col}}
\newcommand{\Row}{{\sf Row}}
\newcommand{\clO}{\mathcal{O}}
\newcommand{\clLk}{\mathcal{L}^{(\kappa)}}
\allowdisplaybreaks
\begin{document}
\title{Low-Complexity Pareto-Optimal 3D Beamforming for the Full-Dimensional Multi-User Massive MIMO Downlink}
\author{W. Zhu$^{1,2}$, H. D. Tuan$^2$, E. Dutkiewicz$^2$, Y. Fang$^1$, and L. Hanzo$^3$
\thanks{$^1$School of Communication and Information Engineering, Shanghai University, Shanghai 200444, China
(email: wenbozhu@shu.edu.cn, yfang@staff.shu.edu.cn); $^2$School of Electrical and Data Engineering, University of Technology Sydney, Broadway, NSW 2007, Australia (email: wenbo.zhu@student.uts.edu.au, tuan.hoang@uts.edu.au, eryk.dutkiewicz@uts.edu.au);
$^3$School of Electronics and Computer Science, University of Southampton, Southampton, SO17 1BJ, U.K (email: lh@ecs.soton.ac.uk) (Corresponding author: Yong Fang)}
\thanks{H. D. Tuan would like to acknowldege the financial support  by the Australian Research Council's Discovery Projects under Grant DP190102501}
\thanks{Y. Fang would like to acknowledge the financial support of the National Natural Science Foundation of China under Grant 61673253}
\thanks{L. Hanzo would like to acknowledge the financial support of the Engineering and Physical Sciences Research Council projects EP/W016605/1 and EP/X01228X/1 as well as of the European Research Council's Advanced Fellow Grant QuantCom (Grant No. 789028)}
}
\date{}
\maketitle
\begin{abstract}
Full-dimensional (FD) multi-user  massive multiple-input multiple output (m-MIMO) systems employ large two-dimensional (2D)
rectangular antenna arrays to control both the azimuth and elevation angles of
signal transmission. We introduce the sum of two outer products  of the azimuth and elevation beamforming vectors having moderate dimensions as a new class of FD beamforming. We show that this low-complexity class
is capable of outperforming 2D beamforming relying on the single outer product of the azimuth and elevation beamforming vectors. It is also capable of performing close to its FD counterpart  of massive dimensions
in terms of either the users' minimum rate or their geometric mean rate (GM-rate), or
sum rate (SR). Furthermore, we also show that even FD beamforming may be outperformed by
our outer product-based improper Gaussian signaling solution. Explicitly, our design is based on
low-complexity algorithms relying on convex problems of moderate dimensions for max-min rate optimization
or on closed-form expressions for GM-rate and SR maximization.
\end{abstract}
\begin{IEEEkeywords}
Full-dimensional (FD) massive MIMO, FD beamforming, improper Gaussian signaling, outer products, rank-two matrix optimization, multi-objective optimization, max-min rate optimization, geometric mean maximization, sum rate maximization
\end{IEEEkeywords}
\section{Introduction}
Full-dimensional (FD) massive multi-input multi-output (m-MIMO) schemes \cite{Nametal13,Kimetal14,Jietal17,Monetal15}
relying on a large two-dimensional (2D) uniformly spaced rectangular antenna array (URA)
have emerged as a practical m-MIMO implementation.
The design of FD m-MIMO systems in terms of antenna tilts
has been considered e.g. in \cite{Lietal16tvt,Liuetal17, NKDA18,NKA19,LLQJ20} and in the references therein.

As the degree of freedom in FD m-MIMO systems is reduced by the channels' spatial
correlations, it is more practical to serve a large number of single-antenna users than a small number of
multiple-antenna users.
The beamforming design of optimizing the users' rate in m-MIMO systems is computationally complex even for the popular zero-forcing (ZF) or regularized zero-forcing (RZF) or conjugate beamforming (CB) scenarios, since it must rely on iteratively solving convex problems of high dimensions \cite{NTDP19,Yuetaltvt20,Nguetal21}. For FD systems, the most popular 2D beamformer relies on beamforming matrices (BMs) represented by the single outer product of the azimuth beamformer (AB) and elevation beamformer \cite{Yinetal14,ALH17,Kanetal17tvt,Wanetal17,Sonetal19}. These ABs and EBs have been separately designed in
\cite{Yinetal14} for the specific scenario of a single user, and in \cite{Sonetal19} for multiple users.
Their joint design has been considered in
\cite{Kanetal17tvt} with the objective of maximizing the sum rate  using semi-definite relaxation, which imposed an extremely high computational complexity.
All these contributions considered FD of moderate dimensions, rather than of massive dimensions, as in m-MIMO schemes. Hence it
is of substantial interest in practice to investigate particular  classes of FD beamforming capable of mitigating the design complexity of
FD m-MIMO systems.

Subject to the sum transmit power constraint, the beamforming design aims for maximizing either the sum rate (SR) or the users’ minimum rate (MR). The SR problem has been widely studied thanks to its tractable formulation, which is based on iterating upon evaluating low-complexity closed-form expressions. However, it leads to assigning a large fraction of the total SR to a few privileged users having the best channel conditions, leaving only low rates for all other users and thus having a potentially zero MR. Maximizing the MR may only be achieved at the cost of degrading the total SR, and it is also computationally challenging, since its computation must be based on iterating by evaluating large-scale convex problems \cite{TTN16,Naetal17tsp,Yuetaljsac20} to handle the non-differentiable objective function and thus it is not practical for m-MIMO systems. One can potentially balance the SR and MR by maximizing the SR subject to a constraint on MR, but this is computationally challenging. Our previous papers \cite{Yuetaltwc22,Zhuetal22tvt,Nasetal22tvt,Nasetal22tcom} have shown that the geometric mean of the users’ rates (GM-rate) is a beneficial objective function, since its maximization leads to a similar rate for all users without having to impose MR constraints, while still maintaining a good SR. Hence the solution becomes reminiscent of a Pareto optimal solution constructed for optimizing the twice-objective SR and MR problem. As strong duality holds under very mild conditions for single-constrained optimization \cite{TT13}, \cite[Chaper 10.2]{Tuybook}, the GM-rate problem considered leads itself to convenient tractable computation, despite the fact that the GM-rate objective function is nonconcave (but differentiable).

Recent studies such as \cite{HJU13,Zeetal13,LAV16,NTDP19spl,Tuetal19,Yuetaljsac20,Yuetaltwc22} and the references therein have shown that  improper Gaussian signaling (IGS) -- which transmits improper Gaussian signals instead of the conventional proper Gaussian signals -- is capable of increasing the users’ rate. However, the beamforming design problem of IGS is much more complex than that under conventional proper Guassian signaling, because each information source intended for each user is beamformed by relying on a pair of vectors -- rather than on a single one -- for which the user rate is defined by a log-determinant function.

Against the above background, this paper is the first one to consider the multi-user beamforming design of FD m-MIMO systems.
The contributions of this paper are three-fold:
\begin{itemize}
\item We define the sum of a few outer products of the ABs and EBs having moderate dimensions as a new
FD beamforming structure, which includes both the 2D and FD beamforming classes as a particular case. More importantly, it is shown that
the sum of two outer products is capable of outperforming 2D beamforming. It also performs similarly to the FD class in terms of the users' MR, or GM-rate, or  SR, despite its lower design complexity. Therefore, the sum-of-two-outer-product may be deemed to be the optimal structure of FD beamforming;
\item We develop a new class of  improper Gaussian signaling, which is based on sum-pairs of the outer products, which is shown
to outperform the FD class;
\item We develop low-complexity algorithms for designing all the beamformers, which are based on convex problems of moderate dimension
for iteratively improving the MR in MR maximization. However, MR maximization results in low SR.
\item Furthermore, we develop closed-form expressions for
iteratively improving the GM-rate and SR. However, SR maximization results in zero rates for the users having low channel quality and thus it is not suitable for providing a fair service for all users.
Our simulations reveal that GM-rate maximization is computationally attractive and it strikes a compelling trade-off between conflicted MR and SR for Pareto optimization.  
\end{itemize}
The contributions of this work are boldly and explicitly contrasted to the literature in Table \ref{tab:CompaIRSon}.

\begin{table*}[!htb]
 \centering
\caption{ Boldly contrasting our novel contributions to the related literature.}
%\begin{center}
  \begin{tabular}{|l|c|c|c|c|c|c|c|c|}
  \hline
  \backslashbox{Contents}{Literature} & \textbf{This work}&\cite{Lietal16tvt,LLQJ20}&\cite{NKDA18,NKA19} & \cite{Kanetal17tvt}&
  \cite{Wanetal17} &\cite{Liuetal17,Sonetal19} \\
  \hline
  FD m-MIMO&$\surd$& & & & $\surd$ &  \\ \hline
  FD beamforming &$\surd$&$\surd$& & & &\\
  \hline
 2D beamforming &$\surd$ &   & $\surd$& $\surd$ & $\surd$ &$\surd$\\
  \hline
 New low-complex beamforming &$\surd$ &  & &  & & \\
  \hline
 Heuristic methods & & $\surd$  &$\surd$ & $\surd$ & $\surd$ &$\surd$\\
  \hline
 Low computational complexity &$\surd$ & $\surd$  & &  &  &$\surd$\\
  \hline
 High computational complexity& &   &$\surd$ &$\surd$  & $\surd$&\\
 \hline
 Pareto optimization &$\surd$ &  & &  & & \\
  \hline
  \end{tabular}
%\end{center}
\label{tab:CompaIRSon}
\end{table*}

The paper is organized as follows. Section II proposes a new class of FD beamforming, whose computational solution is developed
in Section III. Section IV introduces a new class of improper Gaussian signaling, which uses pairs of the sums of outer products, together with a supporting computational solution. Our numerical examples are given in Section V, while Section VI concludes the paper.

{\it Notation.} Only the vector/matrix variables are printed in boldface; $I_N$ is the identity matrix of size $N\times N$. $[X]^2$ is $XX^H$, $\la X,Y\ra=\mbox{trace}(X^HY)$, which is the dot product of
the matrices $X$ and $Y$.
We also write $\la X\ra=\mbox{trace}(X)$ for notational simplicity. Accordingly, the Frobenius norm of $X$ is defined by $||X||=\sqrt{\la [X]^2\ra}$.
Furthermore, $X\succeq 0$ means that the matrix $X$
is positive semi-definite. ${\sf vect}(.)$ arranges the matrix
into a vector by stacking its columns, while $\Col(x_k)_{k\in\clK}$ arranges
the columns $x_k$, $k\in\clK\triangleq \{1,\dots, K\}$  in the single column
$\begin{bmatrix}x_1^T&\dots&x_K^T\end{bmatrix}^T$,
and $\Row(x_k)_{k\in\clK}$ arranges the rows $x_k$, $k\in\clK$,  in the single row $\begin{bmatrix}x_1&\dots&x_K\end{bmatrix}^T$.
Lastly, let us denote the set of circular Gaussian random variables having zero means and variance $a$  by ${\cal C}(0,a)$.

The following inequalities, which were derived in \cite{TTN16}, are frequently used in our theoretical development:
\begin{equation}\label{inv2}
	\ln\left(1+\frac{|v|^2}{y+\sigma}\right)\geq \alpha(\bar{v},\bar{y})+
	2\frac{\Re\{\bar{v}^*v\}}{\bar{y}+\sigma}
	-\psi(\bar{v},\bar{y})(|v|^2+y),
\end{equation}
with
\begin{equation}\label{alpha1}
\begin{array}{c}
\alpha(\bar{v},\bar{y})\triangleq
\ln\left(1+\frac{|\bar{v}|^2}{\bar{y}}\right)-\frac{|\bar{v}|^2}{\bar{y}}-\sigma\psi(\bar{v},\bar{y}),\\
0<\psi(\bar{v},\bar{y})\triangleq \frac{|\bar{v}|^2}{(\bar{y}+\sigma)(|\bar{v}|^2+\bar{y}+\sigma)}
\end{array}
\end{equation}
for all $v\in\mathbb{C}$, $y\geq 0$, $\sigma>0$, and $\bar{v}\in\mathbb{C}$, $\bar{y}\geq 0$, and
\begin{eqnarray}\label{ine2}
	\ln\left|I_2+[V]^2(Y+\sigma I_2)^{-1}\right| \geq\alpha(\bar{V},\bar{Y})\nonumber\\
	+2\Re\{\la (\bar{Y}+\sigma I_2)^{-1}\bar{V}, V\ra\} -\la \Psi(\bar{V},\bar{Y}),[V]^2+Y\ra,
\end{eqnarray}
with
\begin{equation}\label{alpha2}
\begin{array}{c}
\alpha(\bar{V},\bar{Y})\triangleq \ln\left|I_2+[\bar{V}]^2(\bar{Y}+\sigma I_2)^{-1}\right|\\
	-\la (\bar{Y}+\sigma I_2)^{-1},[\bar{V}]^2\ra -\sigma \la \Psi(\bar{V},\bar{Y})\ra,\\
0\preceq \Psi(\bar{V},\bar{Y})\triangleq (\bar{Y}+\sigma I_2)^{-1}-([\bar{V}]^2+\bar{Y}+\sigma I_2)^{-1}
\end{array}
\end{equation}
for all matrices  $V$, $Y\succ 0$, $\bar{V}$, and $\bar{Y}\succ 0$ of size $2\times 2$, and $\sigma>0$.  Considering both sides of (\ref{inv2}) and (\ref{ine2}) as functions of $(v,y)$ and $(V,Y)$, which match each other at $(\bar{v},\bar{y})$
and $(\bar{V},\bar{Y})$, the functions in the right hand side (RHS)  provide tight minorants of their counterparts in the left hand side (LHS) \cite{Tuybook}.
\section{Low-complexity structured beamforming}
We consider a network  of a base station (BS) equipped with a $M\times M$-URA
to serve $K$ single-antenna downlink users (UEs), which are indexed by $k\in\clK\triangleq \{1,\dots, K\}$.
Let $H_k\in\mathbb{C}^{M\times M}$ represent the channel matrix spanning from the BS' URA to UE $k$, i.e. each of its entry $H_k(m,n)$
characterizes the link spanning from the $(m,n)$-th antenna to UE $k$.

Let $s_k\in \clC(0,1)$ be the information symbol intended for UE $k\in\clK$, which is  beamformed by the matrix $\bW_k\in \mathbb{C}^{M\times M}$ for the BS's downlink (DL) transmission, i.e. the $(m,n)$-th antenna transmits the signal $\sum_{k=1}^K\bW_k(m,n)s_k$. The DL signal received at UE $k$ is given by
\begin{eqnarray}
	y_k&=&\sum_{k'=1}^K\la H_k^T\bW_{k'}\ra s_{k'}+n_k,\label{td6e}
\end{eqnarray}
where $n_k\in\clC(0,\sigma)$ is  the noise, which incorporates both the background noise and other uncertainties, such as the channel estimation error \cite{RTN12}. The robust design relying on imperfect channel state information, which explicitly incorporates the channel estimation error into the optimization formulation is beyond the scope of this paper.
Based on the following simplified rank-one model of $H_k$:
\begin{equation}\label{hst1}
H_k=h_k^e(h_k^a)^T,
\end{equation}
with $h_k^a\in\mathbb{C}^{M}$ representing azimuth span while $h_k^e\in\mathbb{C}^{M}$
representing the elevation span, the following $2$D beamformer of rank one has been proposed in \cite{Yinetal14,ALH17,Kanetal17tvt,Wanetal17,Sonetal19}:
\begin{equation}\label{hst2}
\bW_k=\bw^e_{1,k}(\bw^a_{1,k})^T,
\end{equation}
where $\bw^a_{1,k}\in\mathbb{C}^{M}$ and $\bw^e_{1,k}\in\mathbb{C}^{M}$ are termed as the AB and EB,
respectively. In this paper, we propose the following new structure of $\bW_k\in\mathbb{C}^{M\times M}$ of rank $Q$
\begin{equation}\label{bea1}
\bW_k=\sum_{q=1}^Q\bw^e_{q,k}(\bw^a_{q,k})^T,
\end{equation}
with
\begin{equation}\label{td5}
\bw^{\chi}_{q,k}\in\mathbb{C}^{M}, \chi\in\{a,e\}, q\in\clQ\triangleq \{1,\dots, Q\}.
\end{equation}
In other words, each beamforming matrix (BM) $\bW_k\in\mathbb{C}^{M\times M}$ is a sum of $Q$ outer products $\bw^e_{q,k}(\bw^a_{q,k})^T$, $q\in\clQ$, each of which is termed as a 2D BM. As expected, the
$2D$ beamformer (\ref{hst2}) is a particular case of (\ref{bea1}) for $Q=1$, while the unstructured FD $\bW_k\in\mathbb{C}^{M\times M}$ is also a
particular case of (\ref{bea1}) for $Q=M$. Note that each
BM $\bW_k$ in (\ref{bea1}) is characterized by $2QM$ decision variables, while each $2D$ BM in (\ref{hst2}) is characterized by $2M$ decision variables. Looking ahead, in the simulations we will show that the BM $\bW_k$ defined by (\ref{bea1}) for $Q=2$ already performs similarly well to the FD ones, even though the latter are characterized by $M^2$ decision variables.

With $\bW_k$ defined by (\ref{bea1}), equation (\ref{td6e}) of the DL signal received at UE $k$ becomes
\begin{equation}\label{tra2}
y_k=\sum_{k'=1}^K\left(\sum_{q=1}^Q(\bw^e_{q,k'})^TH_k\bw^a_{q,k'}\right)s_{k'}+n_k.
\end{equation}
In what follows, we will use the notations
\begin{equation}\label{td7a}
\bw^{\chi}_{k}\triangleq \Col(\bw^{\chi}_{q,k})_{q\in\clQ}\in \mathbb{C}^{QM}, \chi\in\{a,e\},
\end{equation}
and then $\bw^a\triangleq \{\bw^a_{k}, k\in\clK \}$, and $\bw^e\triangleq \{\bw^e_{k}, k\in\clK \}$, and
$\bw\triangleq \{\bw^a,\bw^e\}$.
The rate at UE $k$ is
\begin{equation}\label{tra3}
r_k(\bw^a,\bw^e)\triangleq \ln\left[1+ g_k(\bw^a,\bw^e) \right],
\end{equation}
where we have
\begin{equation}\label{sinr1}
	g_k(\bw^a,\bw^e)\triangleq \frac{|\sum_{q=1}^Q(\bw_{q,k}^e)^TH_k\bw^a_{q,k}|^2}{\sum_{k'\in\clK\setminus\{k\} }|\sum_{q=1}^Q(\bw_{q,k'}^e)^T H_k\bw^a_{q,k'}|^2+\sigma}.
\end{equation}

Given the power budget $P$, we are interested in the following three problems of
sum-power constrained Pareto beamforming optimization:
\begin{subequations}\label{stra4}
	\begin{eqnarray}
		\max_{\bw=(\bw^a,\bw^e)}\ \varphi_{SR}(\bw^a,\bw^e)\triangleq \sum_{k=1}^Kr_k(\bw^a,\bw^e)\label{stra4a}\\
		\mbox{s.t.}\quad \sum_{k=1}^K||\sum_{q=1}^Q\bw^a_{q,k}(\bw^e_{q,k})^T||^2\leq P,\label{tra4b}
	\end{eqnarray}
\end{subequations}
for  maximizing  the SR as well as
\begin{eqnarray}\label{mtra4}
\max_{\bw=(\bw^a,\bw^e)}\ \varphi_{MR}(\bw^a,\bw^e)\triangleq \min_{k=1, \dots, K}r_k(\bw^a,\bw^e)\nonumber\\ \mbox{s.t.}\quad (\ref{tra4b}),
\end{eqnarray}	
for maximizing  the MR, and
\begin{eqnarray}\label{tra4}
\max_{\bw=(\bw^a,\bw^e)}\ \varphi_{GM}(\bw^a,\bw^e)\triangleq \left(\prod_{k=1}^Kr_k(\bw^a,\bw^e)\right)^{1/K}\nonumber\\ \mbox{s.t.}\quad
(\ref{tra4b}),
\end{eqnarray}
for maximizing the GM-rate. We will compare the SR, MR, and GM-rate performances attained by the
beamforming class (\ref{bea1})  to that by their FD counterparts $\bW\triangleq \{\bW_k\in\mathbb{C}^{M\times M}, k\in\clK\}$:
\begin{subequations}\label{sfdb3}
	\begin{eqnarray}
		\max_{\bW}\ f_{SR}(\bW)\triangleq \sum_{k=1}^Kr_k(\bW)\label{sfdb3a}\\
		\mbox{s.t.}\quad \sum_{k=1}^K||\bW_k||^2\leq P,\label{sfdb3b}
	\end{eqnarray}
\end{subequations}
and
\begin{equation}\label{mfdb3}
\max_{\bW}\ f_{MR}(\bW)\triangleq  \min_{k=1, \dots, K}r_k(\bW)\quad\mbox{s.t.}\quad (\ref{sfdb3b}),
\end{equation}
and
\begin{equation}\label{fdb3}
	\max_{\bW}\ f_{GM}(\bW)\triangleq \left(\prod_{k=1}^Kr_k(\bW)\right)^{1/K}\quad\mbox{s.t.}\quad  (\ref{sfdb3b}),
\end{equation}
where
\begin{equation}r_k(\bW)=\ln\left(1+\frac{|\la H_k,\bW_k\ra|^2}{\sum_{k'\in\clK\setminus\{k\}}|\la H_k,\bW_{k'}\ra|^2+\sigma}\right).
	\end{equation}
As aforementioned, the SR problem (\ref{sfdb3}) is computationally
tractable but it leads to very low rates for  users having less favorable 
channel conditions and thus zero MR.  The MR problem (\ref{mfdb3}) 
is computationally difficult, since its computation  must be based on iterating large-scale convex problems having the computational complexity order of
 ${\cal O}(K^3M^6)$ \cite{TTN16,Naetal17tsp,Yuetaljsac20}   and
 thus it is not practical for FD m-MIMO systems. One can overcome the zero-rate issue of the SR problem (\ref{sfdb3}) by enforcing the additional MR constraint $f_{MR}(\bW)\geq r_{\min}$   for a given $r_{\min}$ to  solve the problem
 \begin{equation}\label{mixed}
 \max_{\bW} f_{SR}(\bW)\quad\mbox{s.t.}\quad (\ref{sfdb3b}), f_{MR}(\bW)\geq r_{\min},
 \end{equation}
 which is however even more computationally intractable than the MR problem (\ref{mfdb3}). This is because the MR constraint in the former is nonconvex and
 thus computationally intractable,  while  the single sum-power constraint 
in the latter is convex and thus it is computationally tractable.  It has been shown in  
our previous papers \cite{Yuetaltwc22,Zhuetal22tvt,Nasetal22tvt,Nasetal22tcom} 
that maximizing  the GM-rate objective function $f_{GM}(\bW)$  leads to  similar rates for all users without 
enforcing the  MR constraint  $f_{MR}(\bW)\geq r_{\min}$, while still maintaining a good SR. Hence again, the resultant solution is reminiscent of a Pareto optimal solution constructed for optimizing the twice-objective SR and MR problem.   Importantly,  we will show that the computational complexity of maximizing the GM-rate objective function is appealingly low, because it is based on iterating by evaluating closed-form expressions.
Another distinct hallmark of maximizing the GM-rate objective in (\ref{fdb3}) that will be highlighted by simulations is the resultant similar  transmit powers at the antennas. This allows us to circumvent the per-antenna transmit power constraints that were deemed important in the m-MIMO implementation of \cite{MLYN16}.
In other words, problem (\ref{tra4}) of maximizing the GM-rate subject to a single sum-power constraint (\ref{tra4b}) provides a refreshingly new approach to m-MIMO designs capable of maintaining similar user rates, despite using similar DL transmit powers at each antenna.

The next section is devoted to the computational solution of the problems (\ref{stra4})-(\ref{tra4}) and their FD counterparts (\ref{sfdb3})-(\ref{fdb3}).
\section{Optimization algorithms}
The first two subsections propose algorithms for computing (\ref{stra4})-(\ref{tra4}), while the last subsection proposes algorithms for
computing (\ref{sfdb3})-(\ref{fdb3}).

In what follows, we use a feasible initialization of  $(w^{a,(0)},w^{e,(0)})$ for (\ref{tra4b}) and then $\wk\triangleq (\wka,\wke)$
denotes a feasible point of (\ref{tra4b}) that is found from the $(\kappa-1)$-st iteration for $\kappa=1, 2,\dots$. The $\kappa$-th iteration is used for generating $\wko=(\wkoa,\wkoe)$ as follows.
\subsection{MR optimization algorithm}
\subsubsection{AB alternating optimization}
With $\bw^e$ held fixed at $\wke$, the constraint (\ref{tra4b}) becomes
\begin{equation}
\sum_{k=1}^K \la \Qek_k,[\bw^a_k]^2\ra\leq P,\label{taa5b}
\end{equation}
where
\begin{equation}\label{qek}
0 \!\preceq\! \Qek_k \!\triangleq\! \begin{bmatrix}\la \wke_{q,k},\wke_{q',k}\ra I_{M}\end{bmatrix}_{(q,q')\in \clQ\times \clQ}\in\mathbb{C}^{(QM)\times (QM)}.
\end{equation}
For $(k',k)\in \clK\times\clK$, we define
\begin{equation}\label{taa2}
\hke_{k',k}\triangleq \Row((\wke_{q,k'})^TH_k)_{q\in\clQ}\in \mathbb{C}^{1\times (QM)},
\end{equation}
to write
\begin{align}\label{taa3}
r_k(\bw^a,\wke)&=r^a_k(\bw^a)\nonumber\\
&\triangleq\ln\left(1+\frac{|\hke_{k,k}\bw^a_{k}|^2}{\sum_{k'\in\clK\setminus\{ k\}}|\hke_{k',k}\bw^a_{k'}|^2+\sigma}   \right).
\end{align}
To seek an AB $\wkoa$ so that
\begin{equation}\label{taa1}
\varphi_{MR}(\wka,\wke)<\varphi_{MR}(\wkoa,\wke),
\end{equation}
we consider the following problem:
\begin{equation}\label{taa5}
\max_{\bw^a} \min_{k=1, \dots, K} r^a_k(\bw^a)\quad
\mbox{s.t.}\quad (\ref{taa5b}).
\end{equation}
For $(\bar{v}_k,\bar{y}_k)\triangleq (\hke_{k,k}\wka_{k}, \sum_{k'\in\clK\setminus\{k\}}|\hke_{k',k}\wka_{k'}|^2)$,
and accordingly $\alpha(\bar{v}_k,\bar{y}_k)$ and $\psi(\bar{v}_k,\bar{y}_k)$ defined  by (\ref{alpha1}),
applying the inequality (\ref{inv2}) yields the following tight minorant of $r^a_k(\bw^a)$ at $\wka$:
\begin{equation}
	\rka_k(\bw^a)\triangleq \aka_k+2\Re\{\bka_k\bw^a_k\}-\cka_k\sum_{k'=1}^K|\hke_{k',k}\bw^a_{k'}|^2,\label{taa7}
\end{equation}
for
\begin{equation}\label{taa8}
\aka_k\triangleq \alpha(\bar{v}_k, \bar{y}_k),
\bka_k\triangleq  \frac{\bar{v}^*_k}{\bar{y}_k+\sigma}\hke_{k,k},
\cka_k\triangleq \psi(\bar{v}_k,\bar{y}_k).
\end{equation}
Thus, we generate $\wkoa$ as the optimal solution of the following convex quadratic problem of tight
minorant maximization:
\begin{equation}\label{taa12m}
\max_{\bw^a}\min_{k=1,\dots, K}\rka_k(\bw^a)\quad\mbox{s.t.}\quad
(\ref{taa5b}).
\end{equation}
As $\wka$ and $\wkoa$ constitute a feasible point and the optimal solution of (\ref{taa12m}), we have (\ref{taa1})
as long as $\varphi_{MR}(\wkoa,\wke)\neq \varphi_{MR}(\wka,\wke)$.
\subsubsection{EB alternating optimization}
With $\bw^a$ held fixed at $\wkoa$, the constraint (\ref{tra4b}) becomes
\begin{equation}
\sum_{k=1}^K \la \Qak_k,[\bw^e_{k}]^2\ra\leq P,\label{tea5b}
\end{equation}
with
\begin{eqnarray}\label{qak}
0\preceq \Qak_k\triangleq \begin{bmatrix}\la \wkoa_{q,k},\wkoa_{q',k}\ra I_{M}
\end{bmatrix}_{(q,q')\in\clQ\times \clQ}\nonumber\\ \in\mathbb{C}^{(QM)\times (QM)}.
\end{eqnarray}
Upon defining
\begin{equation}\label{tea2}
\hka_{k',k}\triangleq \Row((\wkoa_{q,k'})^TH_k^T)_{q\in\clQ}\in\mathbb{C}^{1\times (QM)},
\end{equation}
we may write
\begin{align}\label{tea3}
r_k(\wkoa,\bw^e)&=r^e_k(\bw^a)\nonumber\\
&\triangleq \ln\left(1+\frac{|\hka_{k,k}\bw^e_{k}|^2}{\sum_{k'\in\clK\setminus\{ k\}}|\hka_{k',k}\bw^e_{k'}|^2+\sigma}   \right).
\end{align}
Thus, to seek an  EB $\wkoe$ so that
\begin{equation}\label{tea1}
\varphi_{MR}(\wkoa,\wke)<\varphi_{MR}(\wkoa,\wkoe),
\end{equation}
we consider the following problem:
\begin{equation}\label{tea5}
\max_{\bw^e} \min_{k=1, \dots, K}r^e_k(\bw^e)\quad
\mbox{s.t.}\quad (\ref{tea5b}).
\end{equation}
For $(\bar{v}_k,\bar{y}_k)\triangleq (\hka_{k,k}\wke_{k}, \sum_{k'\in\clK\setminus\{k\}}|\hka_{k',k}\wke_{k'}|^2)$ and
according to $\alpha(\bar{v}_k,\bar{y}_k)$ and $\psi(\bar{v}_k,\bar{y}_k)$ defined by (\ref{alpha1}),
applying the inequality (\ref{inv2})  yields the following tight concave quadratic minorant of the function $r^e_k(\bw^e)$ at $\wke$:
\begin{equation}
\rke_k(\bw^e)\triangleq \ake_k+2\Re\{\bke_k\bw^e_k\}-\cke_k\sum_{k'=1}^K|\hka_{k',k}\bw^e_{k'}|^2,\label{tea7}
\end{equation}
for
\begin{equation}\label{tea8}
\ake_k\triangleq \alpha(\bar{v}_k,\bar{y}_k),
\bke_k\triangleq \ds \frac{\bar{v}^*_k}{\bar{y}_k+\sigma}\hka_{k,k},
\cke_k\triangleq   \psi(\bar{v}_k,\bar{y}_k).
\end{equation}
Thus, we generate $\wkoe$ as the optimal solution of the following  convex quadratic problem of tight minorant maximization of (\ref{tea5}):
\begin{equation}\label{tea12m}
\max_{\bw^e}\min_{k=1,\dots, K}\rke_k(\bw^e)
\quad\mbox{s.t.}\quad (\ref{tea5b}).
\end{equation}
Similar to (\ref{taa1}),  we have (\ref{tea1})
as far as $\varphi_{MR}(\wkoa,\wkoe)\neq \varphi_{MR}(\wkoa,\wke)$.
\subsubsection{The algorithm and its computational complexity}
Algorithm \ref{alg4m} provides the pseudo-code for the proposed computational procedure.
The sequence $\{(\wka,\wke)\}$ is composed of improved feasible points
of problem (\ref{mtra4}) and thus it converges to a point $(\bar{w}^a,\bar{w}^e)$
according to Cauchy's theorem, which satisfies the first-order optimality condition
with $\bw^e$ ($\bw^a$, resp.) held fixed at $\bar{w}^e$ ($\bar{w}^a$, resp.). The computational complexity of each of its iteration is on the order of
\begin{equation}\label{complex1}
\clO(K^3Q^3M^3),
\end{equation}
which is the computational complexity of (\ref{taa12m}) and (\ref{tea12m}) \cite{Peaucelle-02-A}. Clearly, this involves 
$KQM$ decision variables.

\subsection{SR and GM-rate maximization algorithms}
With the concave functions $\rka_k$ and $\rke_k$ defined from (\ref{taa7}) and (\ref{tea7}), their GMs $(\prod_{k=1}^K\rka_k)^{1/K}$
and $(\prod_{k=1}^K\rke_k)^{1/K}$ are still concave functions \cite[Prop. 2.7]{Tuybook}. Therefore, like Algorithm \ref{alg4m}, the following
Algorithm \ref{alg4gm} generates a sequence of $\{(\wka,\wke)\}$ such that
\begin{align}\label{chain3}
\varphi_{GM}(\wka,\wke)&< \varphi_{GM}(\wkoa,\wke)\nonumber\\ &<\varphi_{GM}(\wkoa,\wkoe),
\end{align}
which converges to a point $(\bar{w}^a,\bar{w}^e)$ that satisfies the first-order optimality condition
with $\bw^e$ ($\bw^a$, resp.) held fixed at $\bar{w}^e$ ($\bar{w}^a$, resp.). However, the computational complexity of each of its iteration is given by (\ref{complex1}).

Now, following \cite{Yuetaltwc22,Zhuetal22tvt,Nasetal22tvt,Nasetal22tcom} we develop another algorithm for computing (\ref{tra4}), which is based on iterating by evaluating closed-form
expressions and thus it is much more computationally efficient than Algorithm \ref{alg4gm}. More importantly, our simulations
will show that both Algorithms \ref{alg4gm} and \ref{alg4} below perform similarly, so the advantage of the latter is plausible.

Upon defining $\varphi_{GM}(\bw^a,\bw^e)=\Phi_{GM}[r_1(\bw^a,\bw^e), \dots, r_K(\bw^a,\bw^e)]$ in conjunction with 
\[
\Phi_{GM}(r_1,\dots, r_K)\triangleq \left( \prod_{k=1}^K r_k(\bw^a,\bw^e)\right)^{1/K},
\]
it is readily seen that $\varphi_{GM}(\bw^a, \bw^e)$ is composed of the function $\Phi_{GM}(r)$ and of the mapping
$r(\bw^a,\bw^e)\triangleq (r_1(\bw^a,\bw^e), \dots, r_K(\bw^a,\bw^e))$.  
Then the linearized function of $\Phi_{GM}$ at $r(\wka,\wke)$ becomes
\[
\clLk[r(\bw^a,\bw^e)]=\ds\frac{\Phi_{GM}[r(\wka,\wke)]}{K}
\sum_{k=1}^K\frac{r_k(\bw^a,\bw^e)}{r_k(\wka,\wke)}.
\]
Thus, at the $\kappa$-th iteration we aim for solving the following problem:
\[
\max_{\bw=(\bw^a,\bw^e)}\ \clLk[r(\bw^a,\bw^e)]\quad \mbox{s.t.}\quad
(\ref{tra4b}).
\]
As $\Phi_{GM}[r(\wka,\wke)]/K>0$, this problem is equivalent to 
 the following problem of weighted sum rate maximization:
\begin{equation}\label{tra6}
	\max_{\bw}\varphik(\bw^a,\bw^e)\triangleq \sum_{k'=1}^K\gammak_kr_k(\bw^a,\bw^e)\quad\mbox{s.t.}\quad (\ref{tra4b}),
\end{equation}
with the weights $\gammak_k$ updated at the $\kappa$-th iteration according to
\begin{equation}\label{tra5}
	\gammak_k\triangleq \frac{\max_{k'\in\clK}r_{k'}(\wka,\wke)}{r_k(\wka,\wke)}.
\end{equation}

 \subsubsection{AB alternating optimization}
 To seek an AB $\wkoa$ so that
 \begin{equation}\label{taa1g}
 	\varphik(\wkoa,\wke)> \varphik(\wka,\wke).
 \end{equation}
 we consider the following problem:
 \begin{equation}\label{taa5g}
 	\max_{\bw^a}\varphik_a(\bw^a)\triangleq \sum_{k'=1}^K\gammak_k r^a_k(\bw^a)\quad\mbox{s.t.}\quad
 	(\ref{taa5b}).
 \end{equation}
Let $\rka_k(\bw^a)$ be defined from (\ref{taa7}).
A tight concave quadratic minorant of the function $\varphik_a(\bw^a)$ at $\wka$ is
 \begin{align}
 	\tvarphik_a(\bw^a)\triangleq &\sum_{k=1}^K\gammak_k\rka_k(\bw^a)\label{taa9}\\
 	=&\aka+2\sum_{k=1}^K\Re\{\gammak_k\bka_k\bw^a_k\}\nonumber\\
 	&-\sum_{k=1}^K(\bw^a_{k})^H\Cka_k \bw^a_{k},\label{taa10}
 \end{align}
 for
\begin{equation}\label{taa11}
\begin{array}{c}
\aka\triangleq\sum_{k=1}^K\gammak_k\aka_k,\\
\Cka_k\triangleq \sum_{k'=1}^K\gammak_{k'}\cka_{k'}(\hke_{k,k'})^H\hke_{k,k'}.
\end{array}
\end{equation}
 Thus, we generate $\wkoa$ as the optimal solution of the following problem of concave quadratic
 minorant maximization:
 \begin{equation}\label{taa12g}
 	\max_{\bw^a}\tvarphik_a(\bw^a)\quad\mbox{s.t.}\quad (\ref{taa5b}),
 \end{equation}
 which admits the following closed-form solution
 \begin{equation}\label{taa13}
 	\wkoa_k=\begin{cases}\begin{array}{l}(\Cka_k)^{-1}\gammak_k(\bka_k)^H\\ \mbox{if}\quad \ds\sum_{k=1}^K \la \Qek_k,[(\Cka_k)^{-1}\gammak_k(\bka_k)^H]^2\ra\\ \quad\quad \leq P,\cr
 			(\Cka_k+\lambda \Qek_k )^{-1}\gammak_k(\bka_k)^H\\ \mbox{otherwise},
 		\end{array}
 	\end{cases}
 \end{equation}
 where $\lambda>0$ is found by bisection so that
 \begin{equation}\label{taa14}
 	\sum_{k=1}^K\la \Qek_k,[(\Cka_k+\lambda \Qek_k )^{-1}\gammak_k(\bka_k)^H]^2\ra=P.
 \end{equation}
 As $\wka$ and $\wkoa$ constitute a feasible point and the optimal solution of (\ref{taa12g}), we have  (\ref{taa1g})
 as far as $	\varphik(\wkoa,\wke)\neq \varphik(\wka,\wke)$.
 \subsubsection{EB alternating optimization}
 To seek an EB $\wkoe$ so that
 \begin{equation}\label{tea1s}
 	\varphik(\wkoa,\wkoe)> \varphik(\wkoa,\wke).
 \end{equation}
 we consider the following problem:
 \begin{equation}\label{tea5s}
 	\max_{\bw^e}\varphik_e(\bw^e)\triangleq \sum_{k'=1}^K\gammak_kr^e_k(\bw^e)\quad\mbox{s.t.}\quad
 	(\ref{tea5b}).
 \end{equation}
 For	$\rke_k(\bw^e)$ defined from (\ref{tea7}),
 we obtain the following tight concave quadratic minorant of the function $\varphik_e(\bw^e)$:
 \begin{align}
 	\tvarphik_e(\bw^e)\triangleq&\sum_{k=1}^K\gammak_k\rke_k(\bw^e)\label{tea9}\\
 	=&\ake+2\sum_{k=1}^K\Re\{\gammak_k\bke_k\bw^e_k\}\nonumber\\
 	&-\sum_{k=1}^K(\bw^e_{k})^H\Cke_k \bw^e_{k},\label{tea10}
 \end{align}
 for
 \begin{equation}\label{tea11}
 \begin{array}{c}
 	\ake\triangleq\sum_{k=1}^K\gammak_k\ake_k,\\
 	\Cke_k\triangleq \sum_{k'=1}^K\gammak_{k,k'}\cke_{k'}(\hka_{k,k'})^H\hka_{k,k'}.
 \end{array}
 \end{equation}
 Thus, we generate $\wkoe$ as the optimal solution of the problem
 \begin{equation}\label{tea12}
 	\max_{\bw^e}\tvarphik_e(\bw^e)\quad\mbox{s.t.}\quad
 	(\ref{tea5b}),
 \end{equation}
 which admits the closed-form solution of
 \begin{equation}\label{tea13}
 	\wkoe_k=\begin{cases}\begin{array}{l}(\Cke_k)^{-1}\gammak_k(\bke_k)^H\\ \mbox{if}\quad
 			\ds\sum_{k=1}^K\la \Qak_k,[(\Cke_k)^{-1}\gammak_k(\bke_k)^H]^2\ra\\ \quad\quad \leq P,\cr
 			(\Cke_k+\lambda \Qak_k )^{-1}\gammak_k(\bke_k)^H\\ \mbox{otherwise},
 		\end{array}
 	\end{cases}
 \end{equation}
 where $\lambda>0$ is found by bisection so that
 \begin{equation}\label{tea14}
 	\ds\sum_{k=1}^K  \la \Qak_k,[(\Cke_k+\lambda \Qak_k)^{-1}\gammak_k(\bke_k)^H]^2\ra=P.
 \end{equation}
 Similar to (\ref{taa1g}),  we have (\ref{tea1s})
 as far as $	\varphik(\wkoa,\wkoe)\neq \varphik(\wkoa,\wke)$.
 \subsubsection{The algorithm and its computational complexity}
 \begin{algorithm}[!t]
 	\caption{MR  algorithm} \label{alg4m}
 	\begin{algorithmic}[1]
 		\State \textbf{Initialization:} Randomly generate initial $(w^{a,(0)}, w^{e,(0)})$ feasible for (\ref{tra4b}).
 		\State \textbf{Repeat until convergence of the objective function in (\ref{mtra4}):} Generate AB
 		$\wkoa$ by solving the convex quadratic problem  (\ref{taa12m}) and EB $\wkoe$ by solving the convex quadratic problem (\ref{tea12m}). Reset $\kappa\leftarrow \kappa+1$.
 		\State \textbf{Output} $(\wka,\wke)$.
 	\end{algorithmic}
 \end{algorithm}

 \begin{algorithm}[!t]
 	\caption{Convex-solver based GM-rate  algorithm} \label{alg4gm}
 	\begin{algorithmic}[1]
 		\State \textbf{Initialization:} Randomly generate initial $(w^{a,(0)}, w^{e,(0)})$ feasible for (\ref{tra4b}).
 		\State \textbf{Repeat until convergence of the objective function in (\ref{tra4}):} With $\rka_k$  (\ref{taa7}),
 		generate AB $\wkoa$ by solving the convex  problem  $\max_{\bw^a}(\prod_{k=1}^K\rka_k(\bw^a))^{1/K}\quad\mbox{s.t.}\quad
 		(\ref{taa5b})$. With $\rke_k$ defined from (\ref{tea7}),
 		generate EB $\wkoe$ by solving the convex problem $\max_{\bw^e}(\prod_{k=1}^K\rke_k(\bw^e))^{1/K}
 		\quad\mbox{s.t.}\quad (\ref{tea5b})$. Reset $\kappa\leftarrow \kappa+1$.
 		\State \textbf{Output} $(\wka,\wke)$.
 	\end{algorithmic}
 \end{algorithm}

 \begin{algorithm}[!t]
 	\caption{Closed-form expression based GM-rate algorithm} \label{alg4}
 	\begin{algorithmic}[1]
 		\State \textbf{Initialization:} Randomly generate initial $(w^{a,(0)}, w^{e,(0)})$ feasible for (\ref{tra4}).
 		\State \textbf{Repeat until convergence of the objective function in (\ref{tra4}):} Generate AB
 		$\wkoa$ by (\ref{taa13}) and EB $\wkoe$ by (\ref{tea13}). Reset $\kappa\leftarrow \kappa+1$.
 		\State \textbf{Output} $(\wka,\wke)$.
 	\end{algorithmic}
 \end{algorithm}
 Algorithm \ref{alg4} provides the pseudo-code for the proposed computational procedure based on generating
 AB $\wkoa$ and EB $\wkoe$ by (\ref{taa13}) and (\ref{tea13}), respectively. Note that there is no need to
 employ a line search for locating the step size $\lambda^a\in [0,1]$ and $\lambda^e\in [0,1]$ so that $\varphi(\wka+\lambda^a(\wkoa-\wka),\wke)>\varphi(\wka,\wke)$ for
 updating $\wkoa\leftarrow \wka+\lambda^a(\wkoa-\wka)$ and $\varphi(\wkoa,\wke+\lambda^e(\wkoe-\wke))>\varphi(\wkoa,\wke)$ for
 updating $\wkoe\leftarrow \wke+\lambda^e(\wkoe-\wke)$. This is because as it will be shown by our simulations, we always have the full step size of length one ($\lambda^a=\lambda^e=1$) to confirm (\ref{chain3}),
 i.e. the sequence $\{(\wka,\wke)\}$ is composed of improved feasible points
 of problem (\ref{tra4}) and thus it converges to a point $(\bar{w}^a,\bar{w}^e)$
 according to Cauchy's theorem. As (\ref{tra4}) and (\ref{taa5g}) ((\ref{tea5}), resp.)
 share the same first order
 optimality condition, the point $\bar{w}^a$  ($\bar{w}^e$, resp.) satisfies the first-order optimality condition
 with $\bw^e$ ($\bw^a$, resp.) held fixed at $\bar{w}^e$ ($\bar{w}^a$, resp.). Note that the
 computational complexity of each of its iteration is linear in $KQM$.

 Finally, is is plausible that the SR problem (\ref{stra4}) is computed  by implementing Algorithm \ref{alg4} for $\gammak_k\equiv 1$.
\subsection{Baseline FD beamforming}
For $(\bar{v}_k,\bar{y}_k)\triangleq (\la H^T_k\Wk_{k}\ra, \sum_{k'\in\clK\setminus\{k\}}|\la H^T_{k}\Wk_{k'}\ra|^2)$ and
accordingly $\alpha(\bar{v}_k,\bar{y}_k)$ and $\psi(\bar{v}_k,\bar{y}_k)$ defined by (\ref{alpha1}),
we define $\ak_k\triangleq \alpha(\bar{v}_k,\bar{y}_k)$,
$\Bk_k\triangleq \ds \frac{\bar{v}^*_k}{\bar{y}_k+\sigma}H_k$,
and $\ck_k\triangleq   \psi(\bar{v}_k,\bar{y}_k)$. Following the lines of generating $\wkoa$ by  (\ref{taa13}), we can derive Algorithms \ref{alg7m} and \ref{alg7} of computing
(\ref{mfdb3}) and (\ref{fdb3}), which converge to a point satisfying the first-order optimality condition. The computational complexity of each iteration of Algorithm \ref{alg7m} is $\clO(K^3M^6)$, which is extremely high, while that of its counterpart, namely Algorithm \ref{alg7} is linear in $KM^2$.

\begin{algorithm}[!t]
	\caption{Convex solver based FD MR/GM-rate algorithm} \label{alg7m}
	\begin{algorithmic}[1]
	\State \textbf{Initialization:} Take an initial $W^{(0)}_k$ feasible for (\ref{sfdb3b}). Set $\kappa=0$.
		\State \textbf{Repeat until convergence of the objective function in (\ref{mfdb3})/(\ref{fdb3}):}
For (\ref{mfdb3}), solve the following convex problem to generate $\Wko$: $\ds\max_{\bW}\min_{k=1,\dots,K}[\ak_k+2\Re\{\la (\Bk_k)^T\bW_k\ra\}-\ck_k\sum_{k=1}^K|\la H_{k}^T\bW_{k'}\ra|^2]\quad\mbox{s.t.}\quad
(\ref{sfdb3b})$. For (\ref{fdb3}), solve the following convex problem to generate $\Wko$:
\begin{eqnarray}
&&\ds\max_{\bW}\left(\prod_{k=1}^K[\ak_k+2\Re\{\la (\Bk_k)^T\bW_k\ra\}\right. \nonumber\\ &&\left. -\ck_k\sum_{k=1}^K|\la H_{k}^T\bW_{k'}\ra|^2]\right)^{1/K}\quad \mbox{s.t.}\quad(\ref{sfdb3b}).\nonumber
\end{eqnarray}
Reset $\kappa\leftarrow \kappa+1$.
 \State \textbf{Output} $\Wk$.
	\end{algorithmic}
\end{algorithm}
\begin{algorithm}[!t]
	\caption{Closed-form based FD GM-rate/SR algorithm} \label{alg7}
	\begin{algorithmic}[1]
	\State \textbf{Initialization:} Take an initial $W^{(0)}_k$ feasible for (\ref{sfdb3b}). Set $\kappa=0$.
		\State \textbf{Repeat until convergence of the objective function in (\ref{fdb3})/(\ref{sfdb3}):} For $k\in\clK$ define
$\gammak_k\triangleq \max_{k'\in\clK}r_{k'}(\Wk)/r_k(\Wk)$, $\ak\triangleq\sum_{k=1}^K\gammak_k\ak_k$ in the case of (\ref{fdb3}),
or $\gammak_k\equiv 1$ in the case of (\ref{sfdb3}),
 and $\Ck\triangleq \sum_{k=1}^K\gammak_{k}\ck_{k}({\sf vect}(H_{k}))^*{\sf vect}^T(H_{k})$. Generate ${\sf vect}(\Wko)$ by
$(\Ck)^{-1}\gamma_k({\sf vect}(\Bk_k))^*$ if $\ds\sum_{k=1}^K ||(\Ck)^{-1}\gammak_k({\sf vect}(\Bk_k))^*||^2\leq P$.
Otherwise, generate it by $(\Ck+\lambda I_{M^2} )^{-1}\gamma_k({\sf vect}(\Bk_k))^*$,
where $\lambda>0$ is found by bisection so that $\sum_{k=1}^K||(\Ck+\lambda I_{M^2})^{-1}\gammak_k({\sf vect}(\Bk_k))^*||^2=P$.
Reset $\kappa\leftarrow \kappa+1$.
 \State \textbf{Output} $\Wk$.
	\end{algorithmic}
\end{algorithm}

\section{Outer product-based improper Gaussian signaling}
In this section, each $s_k\in\clC(0,1)$ is  beamformed using $\bW_k$ defined from
(\ref{bea1}), while each $s_k^*$ is beamformed by harnessing
\begin{equation}\label{itd5}
\tilde{\bW}_k\triangleq \sum_{q=1}^Q\tbw^e_{q,k}(\tbw^a_{q,k})^T, \tbw^\chi_{q,k}\in\mathbb{C}^{M}, \chi\in\{a,e\}, q\in\clQ,
\end{equation}
to create the transmit signal
\begin{equation}\label{itra1}
X=\sum_{k=1}^K(\bW_ks_{k}+\tilde{\bW}_ks^*_{k}).
\end{equation}
In contrast to the transmit signal $X=\sum_{k=1}^K\bW_ks_k$ in (\ref{td6e}), which is still proper Gaussian ($\mathbb{E}({\sf vect}(X) {\sf vect}^T(X))=0$),
the transmit signal $X$ defined by (\ref{itra1}) is improper Gaussian associated with $\mathbb{E}({\sf vect}(X) {\sf vect}^T(X))\neq 0$. Recent studies such as \cite{HJU13,Zeetal13,LAV16,NTDP19spl,Tuetal19,Yuetaljsac20,Yuetaltwc22} and the references therein have shown that such improper Gaussian signaling (IGS) can manage the multi-user interference more effectively to enhance the users' rate performance.
 However, the design complexity of IGS
of each pair $(\bW_k,\tilde{\bW}_k)$ is characterized by $4QM$ decision variables.

In what follows, together with $\bw^a_{k}$ and $\bw^e_{k}$ defined in (\ref{td7a}), we also define
\begin{align}\label{va}
\tbw^\chi_{k}&\triangleq \Col(\tbw^{\chi}_{q,k})_{q\in\clQ}\in \mathbb{C}^{QM},\nonumber\\
\bv^{\chi}_{k}&\triangleq \begin{bmatrix}\Re\{\bw^\chi_{k}\}\cr
\Im\{\bw^\chi_{k}\}\cr
\Re\{\tbw^\chi_{k}\}\cr
\Im\{\tbw^\chi_{k}\}
\end{bmatrix}\in \mathbb{R}^{4QM}, \chi\in\{a,e\}
\end{align}
so $\bv^\chi_k$  is the  composite real form of $\bw^{\chi}_k$, and then
\begin{equation}\label{var1}
\begin{array}{c}
\tbw^{\chi}\triangleq \{\tbw^{\chi}_{k}, k\in\clK \},
\hbw^{\chi}_k\triangleq \{\bw^{\chi}_k,\tbw^{\chi}_k\},\\ \hbw^{\chi}\triangleq \{\hbw^\chi_k, k\in\clK\},  \chi\in\{a,e\},
\hbw\triangleq \{\hbw^a,\hbw^e\},\\
\bva\triangleq \{\bva_k, k\in\clK \}, \bve\triangleq \{\bve_k, k\in\clK \}.
\end{array}
\end{equation}
Instead of (\ref{tra2}), the signal received at UE $k$ is now formulated as
\begin{align}
y_k=&\sum_{k'=1}^K\left[\left(\sum_{q=1}^Q(\bw^e_{q,k'})^TH_k\bw^a_{q,k'}\right)s_{k'}+ \right.\nonumber\\
&\left. \left(\sum_{q=1}^Q(\tbw^e_{q,k'})^TH_k\tbw^a_{q,k'}\right)s^*_{k'}\right]+n_k\label{itra2}\\
=&\sum_{k'=1}^K\left[h^e_{k',k}(\bw^e_{k'})\bw^a_{k'} s_{k'}+
\tih^e_{k',k}(\tbw^e_{k'})\tbw^a_{k'}s^*_{k'}\right]+n_k\label{itra2a},
\end{align}
with
\begin{equation}\label{itaa2}
\begin{array}{c}
h^e_{k',k}(\bw^e_{k'})\triangleq \Row((\bw^e_{q,k'})^TH_k)_{q\in\clQ}\in \mathbb{C}^{1\times (QM)},\\
\tih^e_{k',k}(\tbw^e_{k'})\triangleq \Row((\tbw^e_{q,k'})^TH_k)_{q\in\clQ}\in \mathbb{C}^{1\times (QM)}.
\end{array}
\end{equation}
The equivalent composite real form of (\ref{itra2a}) is given by (\ref{rc2})-(\ref{rc5}).
\begin{figure*}[t]
\begin{eqnarray}
\begin{bmatrix}\Re\{y_k\}\cr
\Im\{y_k\}\end{bmatrix}&=&\sum_{k'=1}^K\left[\begin{bmatrix}\Re\{h^e_{k',k}(\bw^e_{k'})\}&-\Im\{h^e_{k',k}(\bw^e_{k'})\}\cr
\Im\{h^e_{k',k}(\bw^e_{k'})\}&\Re\{h^e_{k',k}(\bw^e_{k'})\}\end{bmatrix} \begin{bmatrix} \Re\{\bw^a_{k'}\}&-\Im\{\bw^a_{k'}\}\cr
\Im\{\bw^a_{k'}\}&\Re\{\bw^a_{k'}\}
\end{bmatrix}\right.\nonumber\\
&&\left.+\begin{bmatrix}\Re\{\tih^e_{k',k}(\tbw^e_{k'})\}&-\Im\{\tih^e_{k',k}(\tbw^e_{k'})\}\cr
\Im\{\tih^e_{k',k}(\tbw^e_{k'})\}&\Re\{\tih^e_{k',k}(\tbw^e_{k'})\}\end{bmatrix}\begin{bmatrix} \Re\{\tbw^a_{k'}\}&\Im\{\tbw^a_{k'}\}\cr
\Im\{\tbw^a_{k'}\}&-\Re\{\tbw^a_{k'}\}
\end{bmatrix}\right] \begin{bmatrix}\Re\{s_{k'}\}\cr
\Im\{s_{k'}\}\end{bmatrix}+\begin{bmatrix}\Re\{n_k\}\cr
\Im\{n_k\}\end{bmatrix}\label{rc2}\\
&=&\sum_{k'=1}^K\clH^e_{k',k}(\bva_{k'},\hbw^e_{k'})\begin{bmatrix}\Re\{s_{k'}\}\cr
\Im\{s_{k'}\}\end{bmatrix}+\begin{bmatrix}\Re\{n_k\}\cr
\Im\{n_k\}\end{bmatrix}\label{rc3},
\end{eqnarray}
where
\begin{equation}\label{rc4}
\clH^e_{k',k}(\bva_{k'},\hbw^e_{k'})\triangleq \begin{bmatrix}\hh^e_{1,k',k}(\hbw^e_{k'})\bva_{k'}&\hh^e_{2,k',k}(\hbw^e_{k'})\bva_{k'}\cr
\hh^e_{3,k',k}(\hbw^e_{k'})\bva_{k'}&\hh^e_{4,k',k}(\hbw^e_{k'})\bva_{k'}\end{bmatrix}
\end{equation}
and
\begin{equation}\label{rc5}
\begin{array}{c}
\hh^e_{1,k',k}(\hbw^e_{k'})\triangleq \begin{bmatrix}\Re\{h^e_{k',k}(\bw^e_{k'})\}&-\Im\{h^e_{k',k}(\bw^e_{k'})\}& \Re\{\tih^e_{k',k}(\tbw^e_{k'})\}&-\Im\{\tih^e_{k',k}(\tbw^e_{k'})\} \end{bmatrix}\\
\hh^e_{2,k',k}(\hbw^e_{k'})\triangleq \begin{bmatrix}-\Im\{h^e_{k',k}(\bw^e_{k'})\}&-\Re\{h^e_{k',k}(\bw^e_{k'})\}& \Im\{\tih^e_{k',k}(\tbw^e_{k'})\}&\Re\{\tih^e_{k',k}(\tbw^e_{k'})\} \end{bmatrix}\\
\hh^e_{3,k',k}(\hbw^e_{k'})\triangleq \begin{bmatrix}\Im\{h^e_{k',k}(\bw^e_{k'})\}&\Re\{h^e_{k',k}(\bw^e_{k'})\}& \Im\{\tih^e_{k',k}(\tbw^e_{k'})\}&\Re\{\tih^e_{k',k}(\tbw^e_{k'})\} \end{bmatrix}\\
\hh^e_{4,k',k}(\hbw^e_{k'})\triangleq \begin{bmatrix}\Re\{h^e_{k',k}(\bw^e_{k'})\}&-\Im\{h^e_{k',k}(\bw^e_{k'})\}& -\Re\{\tih^e_{k',k}(\tbw^e_{k'})\}&\Im\{\tih^e_{k',k}(\tbw^e_{k'})\} \end{bmatrix}.
\end{array}
\end{equation}
\hrulefill
\vspace{-0.3cm}
\end{figure*}
Thus,  the rate $\hr_k(\hbw^a,\hbw^e)$
 at UE $k$ is given by $(1/2)\hr^a_k(\bva,\hbw^e)$ \cite{CT06} with
\begin{align}\label{rc6}
\hr^a_k(\bva,\hbw^e)\triangleq& \ln\left|I_2+[\clH^e_{k,k}(\hbw^e_k,\bva_{k}) ]^2 \right. \nonumber\\ &\left. \left(\sum_{k'\in\clK\setminus \{k\}}[\clH^e_{k',k}(\hbw^e_{k'},\bva_{k'})]^2+\sigma I_2 \right)^{-1}  \right|.
\end{align}
Analogously, we rewrite the signal received at UE $k$ in (\ref{itra2}) as
\begin{equation}
y_k=\sum_{k'=1}^K\left[h^a_{k',k}(\bw^a_{k'})\bw^e_{k'} s_{k'}+
\tih^a_{k',k}(\tbw^a_{k'})\tbw^e_{k'}s^*_{k'}\right]+n_k\label{itra2e},
\end{equation}
with
\begin{equation}\label{taa2e}
\begin{array}{c}
h^a_{k',k}(\bw^a_{k'})\triangleq \Col((\bw^a_{q,k'})^TH_k^T)_{q\in\clQ}\in \mathbb{C}^{1\times (QM)},\\
\tih^a_{k',k}(\tbw^a_{k'})\triangleq \Col((\tbw^a_{q,k'})^TH_k^T)_{q\in\clQ}\in \mathbb{C}^{1\times (QM)}.
\end{array}
\end{equation}
The equivalent composite real form of (\ref{itra2e}) is
\begin{equation}
\begin{bmatrix}\Re\{y_k\}\cr
\Im\{y_k\}\end{bmatrix}=\sum_{k'=1}^K\clH^a_{k',k}(\hbw^a_{k'},\bve_{k'})\begin{bmatrix}\Re\{s_{k'}\}\cr
\Im\{s_{k'}\}\end{bmatrix}+\begin{bmatrix}\Re\{n_k\}\cr
\Im\{n_k\}\end{bmatrix}\label{rc3e},
\end{equation}
where we have
\begin{equation}\label{rc4e}
\clH^a_{k',k}(\hbw^a_{k'},\bve_{k'})\triangleq \begin{bmatrix}\hh^a_{1,k',k}(\hbw^a_{k'})\bve_{k'}&\hh^a_{2,k',k}(\hbw^a_{k'})\bve_{k'}\cr
\hh^a_{3,k',k}(\hbw^a_{k'})\bve_{k'}&\hh^a_{4,k',k}(\hbw^a_{k'})\bve_{k'}\end{bmatrix}
\end{equation}
and (\ref{rc5e}).
\begin{figure*}[t]
\begin{equation}\label{rc5e}
\begin{array}{c}
\hh^a_{1,k',k}(\hbw^a_{k'})\triangleq \begin{bmatrix}\Re\{h^a_{k',k}(\bw^a_{k'})\}&-\Im\{h^a_{k',k}(\bw^a_{k'})\}& \Re\{\tih^a_{k',k}(\tbw^a_{k'})\}&-\Im\{\tih^a_{k',k}(\tbw^a_{k'})\} \end{bmatrix}\\
\hh^a_{2,k',k}(\hbw^a_{k'})\triangleq \begin{bmatrix}-\Im\{h^a_{k',k}(\bw^a_{k'})\}&-\Re\{h^a_{k',k}(\bw^a_{k'})\}& \Im\{\tih^a_{k',k}(\tbw^a_{k'})\}&\Re\{\tih^a_{k',k}(\tbw^a_{k'})\} \end{bmatrix}\\
\hh^a_{3,k',k}(\hbw^a_{k'})\triangleq \begin{bmatrix}\Im\{h^a_{k',k}(\bw^a_{k'})\}&\Re\{h^a_{k',k}(\bw^a_{k'})\}& \Im\{\tih^a_{k',k}(\tbw^a_{k'})\}&\Re\{\tih^a_{k',k}(\tbw^a_{k'})\} \end{bmatrix}\\
\hh^a_{4,k',k}(\hbw^a_{k'})\triangleq \begin{bmatrix}\Re\{h^a_{k',k}(\bw^a_{k'})\}&-\Im\{h^a_{k',k}(\bw^a_{k'})\}& -\Re\{\tih^a_{k',k}(\tbw^a_{k'})\}&\Im\{\tih^a_{k',k}(\tbw^a_{k'})\} \end{bmatrix}.
\end{array}
\end{equation}
\hrulefill
\vspace{-0.5cm}
\end{figure*}
The rate $\hr_k(\hbw^a,\hbw^e)$
 at UE $k$ is also $(1/2)\hr^e_k(\hbw^a,\bve)$ with
\begin{align}\label{rc6e}
\hr^e_k(\hbw^a,\bve)\triangleq &\ln\left|I_2+[\clH^a_{k,k}(\hbw^a_k,\bve_{k}) ]^2\right. \nonumber\\ &\left. \left(\sum_{k'\in\clK\setminus\{ k\}}[\clH^a_{k',k}(\hbw^e_{k'},\bva_{k'})]^2+\sigma I_2 \right)^{-1}  \right|.
\end{align}
Thus, we have the three problems corresponding to (\ref{stra4}), (\ref{mtra4}), and (\ref{tra4}):
\begin{subequations}\label{sitra4}
\begin{align}
&\max_{\hbw^a,\hbw^e}\ \hvarphi_{SR}(\hbw^a,\hbw^e)\triangleq \sum_{k=1}^K\hr_k(\hbw^a,\hbw^e)\label{sitra4a}\\
\mbox{s.t.}\ \sum_{k=1}^K&\left[||\sum_{q=1}^Q\bw^e_{q,k}(\bw^a_{q,k})^T||^2+||\sum_{q=1}^Q\tbw^e_{q,k}(\tbw^a_{q,k})^2||^2
\right] \leq P,\label{itra4b}
\end{align}
\end{subequations}
and
\begin{equation}\label{mitra4}
\max_{\hbw^a,\hbw^e}\ \hvarphi_{MR}(\hbw^a,\hbw^e)\triangleq \min_{k=1,\dots, K}\hr_k(\hbw^a,\hbw^e)\quad
\mbox{s.t.}\quad (\ref{itra4b}),
\end{equation}
and
\begin{equation}\label{itra4}
\max_{\hbw^a,\hbw^e}\ \hvarphi_{GM}(\hbw^a,\hbw^e)\triangleq \left(\prod_{k=1}^K\hr_k(\hbw^a,\hbw^e)\right)^{1/K}\
\mbox{s.t.}\quad (\ref{itra4b}).
\end{equation}
The first subsection is devoted to computing the MR problem (\ref{mitra4}), while the second subsection is
dedicated to computing the SR problem (\ref{sitra4}) and the GM-rate problem (\ref{itra4}).
In what follows we used a feasible initialization of $\hat{w}^{(0)}\triangleq (\hat{w}^{a,(0)},\hat{w}^{e,(0)})$ for (\ref{itra4b}) and then
$\hwk\triangleq (\hwka,\hwke)$ denotes a feasible point of (\ref{itra4b}) that is found from the $(\kappa-1)$-st iteration.
The $\kappa$-th iteration is used for generating $\hwko\triangleq (\hwkoa,\hwkoe)$ as follows.
\subsection{MR maximization algorithm}
\subsubsection{AB alternating optimization}
We seek an $\hwkoa$ so that
\begin{equation}\label{miaa1}
\hvarphi_{MR}(\hwkoa,\hwke)> \hvarphi_{MR}(\hwka,\hwke).
\end{equation}
Note that we have $\hr_k(\hbw^a,\hwke)=\hrak_k(\bva)$, with
\begin{align}\label{aobj3}
\hrak_k(\bva)\triangleq &\hr^a_k(\bva,\hwke)\nonumber\\ =&\ln\left|I_2+[\clHek_{k,k}(\bva_{k}) ]^2 \right. \nonumber\\ &\left. \left(\sum_{k'\in\clK\setminus\{ k\}}[\clHek_{k',k}(\bva_{k'})]^2+\sigma I_2 \right)^{-1}  \right|,
\end{align}
where
\begin{equation}\label{aobj4}
\clHek_{k',k}(\bva_{k'})\triangleq \begin{bmatrix}\hhek_{1,k',k}\bva_{k'}&\hhek_{2,k',k}\bva_{k'}\cr
\hhek_{3,k',k}\bva_{k'}&\hhek_{4,k',k}\bva_{k'}\end{bmatrix}
\end{equation}
with
\begin{equation}\label{aobj5}
\hhek_{\ell,k',k}\triangleq \hh^e_{\ell,k',k}(\hwke_{k'}), \ell=1, 2, 3, 4
\end{equation}
according to (\ref{rc4})-(\ref{rc5}). Here and after, we use the notation $\vka_k$ and $\vkoa_k$ for the composite
real form of $\hwka_k$ as well as $\hwkoa_k$, and accordingly $\vka\triangleq \{\vka_k, k\in\clK\}$ and $\vkoa\triangleq \{\vkoa_k, k\in\clK\}$.
With $\hbw^e$ held fixed at $\hwke$, the power constraint (\ref{itra4b}) becomes:
\begin{eqnarray}
&&\la \Qek_k,[\bw^a_k]^2\ra + \la \tQek_k,[\tbw^a_k]^2\ra\leq P\label{rc7}\\
&\Leftrightarrow&\la \hQek, [\bva_k]^2\ra\leq P, \label{rc8}
\end{eqnarray}
where $\Qek_k$ is defined from (\ref{qek}), while
\begin{equation}\label{tqek}
0 \!\preceq\! \tQek_k \!\triangleq\! \begin{bmatrix}\la \twke_{q,k},\twke_{q',k}\ra I_{M}
\end{bmatrix}_{(q,q')\in\clQ\times \clQ}\in\mathbb{C}^{(QM)\times (QM)},
\end{equation}
and (\ref{hQe})\footnote{$\Re\{\Qek_k\}$ and  $\Re\{\tQek_k\}$ are symmetric while $\Im\{\Qek_k\}$ and  $\Im\{\tQek_k\}$ are skew-symmetric }.
\begin{figure*}[t]
\begin{equation}\label{hQe}
\hQek_k\triangleq \begin{bmatrix}\Re\{\Qek_k\}&(\Im\{\Qek_k\})^T&0_{(QM)\times (QM)}&0_{(QM)\times (QM)}\cr
\Im\{\Qek_k\}&\Re\{\Qek_k\}&0_{(QM)\times (QM)}&0_{(QM)\times (QM)}\cr
0_{(QM)\times (QM)}&0_{(QM)\times (QM)}&\Re\{\tQek_k\}&(\Im\{\tQek_k\})^T\cr
0_{(QM)\times (QM)}&0_{(QM)\times (QM)}&\Im\{\tQek_k\}&\Re\{\tQek_k\}
\end{bmatrix}.
\end{equation}
\hrulefill
\vspace{-0.5cm}
\end{figure*}

Let $(\bar{V}, \bar{Y})\triangleq (\clHek_{k,k}(\vka_k), \sum_{k'\neq k}^K\clHek_{k',k}[\vka_{k'}]^2)$ and
accordingly $\alpha(\bar{X},\bar{Y})$ and $\Psi(\bar{X},\bar{Y})$ defined by (\ref{alpha2}).
Applying the inequality (\ref{ine2}) yields the following tight minorant of $\hrak_k(\bva)$ given by (\ref{iaa6})-(\ref{iaa8c})\footnote{$\sqrt{\Cka_k}$ is positive definite and so it is symmetric}.
\begin{figure*}[t]
\begin{eqnarray}
\hrhoak_k(\bva)&\triangleq&\aka_k+2\la \Bka_k\clHek_{k,k}(\bva_k)\ra-\la \Cka_k\sum_{k'=1}^K[\clHek_{k',k}(\bva_{k'})]^2\ra\label{iaa6}\\
&=&\aka_k+2\Bka_k(1,1)\hhek_{1,k,k}\bva_{k}+2\Bka_k(1,2)\hhek_{3,k,k}\bva_{k}
+2\Bka_k(2,1)\hhek_{2,k,k}\bva_{k}\nonumber\\
&&+2\Bka_k(2,2)\hhek_{4,k,k}\bva_{k}-
\sum_{k'=1}^K\left|\left|\sqrt{\Cka_k}\clHek_{k',k}(\bva_{k'})\right|\right|^2 \label{iaa6b}\\
&=&\aka_k+2\bka_k\bva_{k}-\sum_{k'=1}^K\left[||(\cka_{k,1}\hhek_{1,k',k}+
\cka_{k,2}\hhek_{3,k',k})\bva_{k'}||^2\right.\nonumber\\
&&+||(\cka_{k,1}\hhek_{2,k',k}+
\cka_{k,2}\hhek_{4,k',k})\bva_{k'}||^2 + ||(\cka_{k,2}\hhek_{1,k',k}+
\cka_{k,3}\hhek_{3,k',k})\bva_{k'}||^2\nonumber\\
&&\left.+||(\cka_{k,2}\hhek_{2,k',k}+
\cka_{k,3}\hhek_{4,k',k})\bva_{k'}||^2  \right]\label{taa6c}\\
&=&\aka_k+2\hbka_k\bva_{k}-\sum_{k'=1}^K\la \Psi^a_{k',k},[\bva_{k'}]^2\ra,\label{iaa6d}
\end{eqnarray}
for
\begin{equation}\label{iaa8}
\begin{array}{c}
\aka_k\triangleq \alpha(\bar{X}, \bar{Y}), 0\preceq \Cka_k\triangleq \Psi(\bar{V},\bar{Y})
\in\mathbb{R}^{2\times 2},  \\
\Bka_k=\begin{bmatrix}\Bka_k(1,1)&\Bka_k(1,2)\cr
\Bka_k(2,1)&\Bka_k(2,2)
\end{bmatrix}
\triangleq \ds (\bar{V}_k)^T(\bar{Y}_k+\sigma I_2)^{-1}\in\mathbb{R}^{2\times 2},
\end{array}
\end{equation}
and
\begin{equation}\label{iaa8b}
\begin{array}{c}
\hbka_k\triangleq \Bka_k(1,1)\hhek_{1,k,k}+\Bka_k(1,2)\hhek_{3,k,k}+\Bka_k(2,1)\hhek_{2,k,k}+\Bka_k(2,2)\hhek_{4,k,k},\\
\begin{bmatrix}\cka_{k,1}&\cka_{k,2}\cr
\cka_{k,2}&\cka_{k,3}\end{bmatrix}\triangleq\sqrt{\Cka_k},
\end{array}
\end{equation}
and
\begin{eqnarray}
\Psi^a_{k',k}&\triangleq& [(\cka_{k,1}\hhek_{1,k',k}+
\cka_{k,2}\hhek_{3,k',k})^T]^2+[(\cka_{k,1}\hhek_{2,k',k}+
\cka_{k,2}\hhek_{4,k',k})^T]^2\nonumber\\
&&+[(\cka_{k,2}\hhek_{1,k',k}+
\cka_{k,3}\hhek_{3,k',k})^T]^2+[(\cka_{k,2}\hhek_{2,k',k}+
\cka_{k,3}\hhek_{4,k',k})^T]^2.\label{iaa8c}
\end{eqnarray}
\hrulefill
\vspace{-0.5cm}
\end{figure*}

We thus solve the following quadratic convex problem of tight minorant maximization of the max-min optimization problem (\ref{mitra4}) with $\hbw^e$ held fixed at $\hwke$:
\begin{equation}\label{mra1}
\max_{\bva}\min_{k=1,\dots, K} \hrhoak_k(\bva)\quad\mbox{s.t.}\quad (\ref{rc7}),
\end{equation}
whose optimal solution $\hwkoa$ verifies (\ref{miaa1}).
\subsubsection{EB alternating optimization}
We seek an EB $\hwkoe$ so that
\begin{equation}\label{mtae1}
\hvarphi_{MR}(\hwkoa,\hwkoe)> \hvarphi_{MR}(\hwkoa,\hwke).
\end{equation}
Note that $\hr_k(\hwkoa,\hbw^e)=\hrek_k(\bve)$ with
\begin{align}\label{eobj3}
\hrek_k(\bve)\triangleq &\hr^e_k(\hwkoa,\bve)\nonumber\\ =&\ln\left|I_2+[\clHak_{k,k}(\bve_{k}) ]^2 \right. \nonumber\\ &\left. \left(\sum_{k'in\clK\setminus \{k\}}^K[\clHak_{k',k}(\bve_{k'})]^2+\sigma I_2 \right)^{-1}  \right|,
\end{align}
where
\begin{equation}\label{eobj4}
\clHak_{k',k}(\bve_{k'})\triangleq \begin{bmatrix}\hhak_{1,k',k}\bve_{k'}&\hhak_{2,k',k}\bve_{k'}\cr
\hhak_{3,k',k}\bve_{k'}&\hhak_{4,k',k}\bve_{k'}\end{bmatrix}
\end{equation}
with
\begin{equation}\label{eobj5}
\hhak_{\ell,k',k}\triangleq \hh^a_{\ell,k',k}(\hwkoa_{k'}), \ell=1, 2, 3, 4
\end{equation}
according to (\ref{rc4e})-(\ref{rc5e}).  Here and after, we use the notation $\vke_k$ and $\vkoe_k$ for the composite
real form of $\hwke_k$ and $\hwkoe_k$, and accordingly $\vke\triangleq \{\vke_k, k\in\clK\}$ and $\vkoe\triangleq \{\vkoe_k, k\in\clK\}$. With $\hbw^a$ held fixed at $\hwkoa$, the power constraint (\ref{itra4b}) becomes:
\begin{eqnarray}
&&\la \Qak_k,[\bw^e_k]^2\ra + \la \tQak_k,[\tbw^4_k]^2\ra\leq P\label{rc7e}\\
&\Leftrightarrow&\la \hQak, [\bve_k]^2\ra\leq P, \label{rc8e}
\end{eqnarray}
where $\Qak_k$ is defined in (\ref{qak}), while
\begin{eqnarray}\label{tqak}
0\preceq \tQak_k\triangleq \begin{bmatrix}\la \twkoa_{q,k},\twkoa_{q',k}\ra I_{M}
\end{bmatrix}_{(q,q')\in\clQ\times \clQ}\nonumber\\ \in\mathbb{C}^{(QM)\times (QM)},
\end{eqnarray}
and (\ref{hQa}).
\begin{figure*}[t]
\begin{equation}\label{hQa}
\hQak_k\triangleq \begin{bmatrix}\Re\{\Qak_k\}&(\Im\{\Qak_k\})^T&0_{(QM)\times (QM)}&0_{(QM)\times (QM)}\cr
\Im\{\Qak_k\}&\Re\{\Qak_k\}&0_{(QM)\times (QM)}&0_{(QM)\times (QM)}\cr
0_{(QM)\times (QM)}&0_{(QM)\times (QM)}&\Re\{\tQak_k\}&(\Im\{\tQak_k\})^T\cr
0_{(QM)\times (QM)}&0_{(QM)\times (QM)}&\Im\{\tQak_k\}&\Re\{\tQak_k\}
\end{bmatrix}.
\end{equation}
\hrulefill
\vspace{-0.5cm}
\end{figure*}

Let $(\bar{V},\bar{Y})\triangleq (\clHak_{k,k}(\vke_k),\sum_{k'\neq k}^K\clHak_{k',k}[\vke_{k'}]^2)$, and
accordingly $\alpha(\bar{V},\bar{Y})$ defined by (\ref{alpha2}).
Like (\ref{iaa6d}), applying the inequality (\ref{ine2}) yields the following tight minorant of $\hrek_k(\bve)$ given by (\ref{tae6d})-(\ref{tae8c}).
\begin{figure*}[t]
\begin{eqnarray}
\hrhoek_k(\bve)\triangleq \ake_k+2\hbke_k\bve_{k}-\sum_{k'=1}^K\la \Psi^e_{k',k},[\bve_{k'}]^2\ra,\label{tae6d}
\end{eqnarray}
for
\begin{equation}\label{tae8}
\begin{array}{c}
\ake_k\triangleq \alpha(\bar{V}, \bar{Y}), 0\preceq \Cke_k\triangleq \Psi(\bar{V},\bar{Y}),\\
\Bke_k=\begin{bmatrix}\Bke_k(1,1)&\Bke_k(1,2)\cr
\Bke_k(2,1)&\Bke_k(2,2)
\end{bmatrix}\triangleq \ds (\bar{V}_k)^T(\bar{Y}_k+\sigma I_2)^{-1}\in\mathbb{R}^{2\times 2},
\end{array}
\end{equation}
and
\begin{equation}\label{tae8b}
\begin{array}{c}
\hbke_k\triangleq \Bke_k(1,1)\hhak_{1,k,k}+\Bke_k(1,2)\hhak_{3,k,k}+\Bke_k(2,1)\hhak_{2,k,k}+\Bke_k(2,2)\hhak_{4,k,k},\\
\begin{bmatrix}\cke_{k,1}&\cke_{k,2}\cr
\cke_{k,2}&\cke_{k,3}\end{bmatrix}\triangleq\sqrt{\Cke_k},
\end{array}
\end{equation}
and
\begin{eqnarray}
\Psi^e_{k',k}&\triangleq& [(\cke_{k,1}\hhak_{1,k',k}+
\cke_{k,2}\hhak_{3,k',k})^T]^2+[(\cke_{k,1}\hhak_{2,k',k}+
\cke_{k,2}\hhak_{4,k',k})^T]^2\nonumber\\
&&+[(\cke_{k,2}\hhak_{1,k',k}+
\cke_{k,3}\hhak_{3,k',k})^T]^2+[(\cke_{k,2}\hhak_{2,k',k}+
\cke_{k,3}\hhak_{4,k',k})^T]^2.\label{tae8c}
\end{eqnarray}
\hrulefill
\vspace{-0.5cm}
\end{figure*}

We thus solve the following quadratic convex problem of tight minorant maximization of the max-min optimization problem (\ref{mitra4}) with $\hbw^a$ held fixed at $\hwkoa$:
\begin{equation}\label{mta1}
\max_{\bve}\min_{k=1,\dots, K} \hrhoek_k(\bve)\quad\mbox{s.t.}\quad (\ref{rc7e}),
\end{equation}
whose optimal solution $\hwkoe$ verifies (\ref{mtae1}).
\subsubsection{The algorithm and its computational complexity}
 Algorithm \ref{mialg4} provides the pseudo-code for the proposed computational procedure based on generating
  $\vkoa$  and  $\vkoe$ by solving (\ref{mra1}) and (\ref{mta1}) of computational complexity on the order of
  \begin{equation}\label{comc}
  \clO(K^3(4QM)^3), 
\end{equation}
which converges like Algorithms \ref{alg4m} and \ref{alg4gm}.

\begin{algorithm}[!t]
	\caption{IGS MR algorithm} \label{mialg4}
	\begin{algorithmic}[1]
			\State \textbf{Initialization:} Randomly generate initial $\hat{w}^{(0)}\triangleq (\hat{w}^{a,(0)}, \hat{w}^{e,(0)})$ feasible for (\ref{itra4b}). Set $\kappa=0$.
			\State \textbf{Repeat until convergence of the objective function in (\ref{mitra4}):}
			Generate the composite real form $\vkoa$ of $\hwkoa$ by solving the convex problem (\ref{mra1}) and the composite real form
			$\vkoe$ of $\hwkoe$ by solving the convex problem (\ref{mta1}). Reset $\kappa\leftarrow \kappa+1$.
			\State \textbf{Output} $\hwk\triangleq (\hwka,\hwke)$.
		\end{algorithmic}
\end{algorithm}

\subsection{SR and GM-rate optimization algorithm}
Similarly to (\ref{tra5}), the $\kappa$-th iteration is based on the following nonconvex problem:
\begin{equation}\label{itra6}
\max_{\hbw^a,\hbw^e}\hvarphik(\hbw^a,\hbw^e)\triangleq \sum_{k=1}^K\hgammak_k\hr_k(\hbw^a,\hbw^e)\quad\mbox{s.t.}\quad (\ref{itra4b}),
\end{equation}
for
\begin{equation}\label{itra5}
\hgammak_k\triangleq \frac{\max_{k'\in\clK}\hr_{k'}(\hwke,\hwka)}{\hr_k(\hwke,\hwka)}, k\in\clK.
\end{equation}
\subsubsection{AB alternating optimization}
We seek an AB  $\hwkoa$ so that
\begin{equation}\label{iaa1}
\hvarphik(\hwkoa,\hwke)> \hvarphik(\hwka,\hwke),
\end{equation}
by considering  the problem
\begin{equation}\label{aobj7}
\max_{\bva} \hvarphiak(\bva)\triangleq \sum_{k=1}^K\hgammak_k\hrak_k(\bva)
\quad\mbox{s.t.}\quad (\ref{rc8})
\end{equation}
Using $\hrhoak_k(\bva)$ defined from (\ref{iaa6d}), which is a tight minorant of  $\hrak_k(\bva)$, we obtain the following
tight minorant of $\hvarphiak(\bva)$:

\begin{align}
\tvarphik_a(\bva)\triangleq&\sum_{k=1}^K\hgammak_k\hrhoak_k(\bva)\label{iaa9}\\
=&\aka+2\sum_{k=1}^K\Re\{\gammak_k\hbka_k\bva_k\}\nonumber\\
&-\sum_{k=1}^K(\bva_{k})^T\Phi^a_k \bva_{k},\label{iaa10}
\end{align}
for
\begin{equation}\label{iaa11}
\aka\triangleq\sum_{k=1}^K\hgammak_k\aka_k, \Phi^a_k\triangleq \sum_{k'=1}^K\hgammak_{k'}\Psi^a_{k,k'}.
\end{equation}
Thus, we generate $\vkoa$ as the optimal solution of the problem for verifying (\ref{iaa1}):
\begin{equation}\label{iaa12}
\max_{\bva}\tvarphik_a(\bva)\triangleq \sum_{k=1}^K\hgammak_k\hrak_k(\bva)\quad\mbox{s.t.}\quad
(\ref{rc8})
\end{equation}
which admits the closed-form solution of
\begin{equation}\label{iaa13}
\vkoa_k=\begin{cases}\begin{array}{l}(\Phi^a_k)^{-1}\hgammak_k(\hbka_k)^T\\ \mbox{if}\quad \ds\sum_{k=1}^K \la \hQek_k,[(\Phi^a_k)^{-1}\hgammak_k(\hbka_k)^T]^2\ra\leq P,\cr
(\Phi^a_k+\lambda \hQek_k)^{-1}\hgammak_k(\hbka_k)^T\\ \mbox{otherwise},
\end{array}
\end{cases}
\end{equation}
where $\lambda>0$ is found by bisection so that
\begin{equation}\label{iaa14}
\sum_{k=1}^K\la \hQek_k,[(\Phi^a_k+\lambda \hQek_k )^{-1}\hgammak_k(\hbka_k)^T]^2\ra=P.
\end{equation}
We then recover $\wkoa$ by using (\ref{va}).
\subsubsection{EB alternating optimization}
We seek an AB  $\hwkoa$ so that
\begin{equation}\label{iae1}
\hvarphik(\hwkoa,\hwkoe)> \hvarphik(\hwkoa,\hwke),
\end{equation}
by considering  the problem
\begin{equation}\label{eobj7}
\max_{\bve} \hvarphiek(\bve)\triangleq \sum_{k=1}^K\hgammak_k\hrek_k(\bva)
\quad\mbox{s.t.}\quad (\ref{rc8e})
\end{equation}
Using $\hrhoek_k(\bve)$ defined from (\ref{tae6d}), which is a tight minorant of  $\hrek_k(\bve)$, we obtain the following
tight minorant of $\hvarphiek(\bve)$:
\begin{align}
\tvarphik_e(\bve)\triangleq&\sum_{k=1}^K\hgammak_k\hrhoek_k(\bve)\label{iaae9}\\
=&\ake+2\sum_{k=1}^K\Re\{\gammak_k\hbke_k\bve_k\}\nonumber\\
&-\sum_{k=1}^K(\bve_{k})^T\Phi^e_k \bve_{k},\label{iae10}
\end{align}
for
\begin{equation}\label{iae11}
\ake\triangleq\sum_{k=1}^K\hgammak_k\ake_k, \Phi^e_k\triangleq \sum_{k'=1}^K\hgammak_{k'}\Psi^e_{k,k'}.
\end{equation}
Thus, we generate $\vkoa$ as the optimal solution of the problem for verifying (\ref{iae1}):
\begin{equation}\label{tae12}
\max_{\bve}\tvarphik_e(\bve)\quad\mbox{s.t.}\quad (\ref{rc8e}),
\end{equation}
which admits the following closed-form solution
\begin{equation}\label{tae13}
\vkoe_k=\begin{cases}\begin{array}{l}(\Phi^e_k)^{-1}\hgammak_k(\hbke_k)^T\\ \mbox{if}\quad \ds\sum_{k=1}^K \la \hQak_k,[(\Phi^e_k)^{-1}\hgammak_k(\hbke_k)^T]^2\ra\leq P,\cr
(\Phi^e_k+\lambda \hQak_k)^{-1}\hgammak_k(\hbke_k)^T\\ \mbox{otherwise},
\end{array}
\end{cases}
\end{equation}
where $\lambda>0$ is found by bisection so that
\begin{equation}\label{tae14}
\sum_{k=1}^K\la \hQak_k,[(\Phi^e_k+\lambda \hQak_k )^{-1}\hgammak_k(\hbke_k)^T]^2\ra=P.
\end{equation}
We then recover $\wkoe$ by using (\ref{va}).
\subsubsection{The algorithm and its computational complexity}
Algorithm \ref{ialg4} provides the pseudo-code for the proposed computational procedure based on generating  $\vkoa$ and  $\vkoe$ by (\ref{iaa13}) and (\ref{tae13}) of linear computational complexity in $K4QM$, which converges similarly to Algorithm \ref{alg4}.

\begin{algorithm}[!t]
	\caption{IGS GM-rate/SR algorithm} \label{ialg4}
	\begin{algorithmic}[1]
	\State \textbf{Initialization:} Randomly generate initial $\hat{w}^{(0)}\triangleq (\hat{w}^{a,(0)}, \hat{w}^{e,(0)})$ feasible for (\ref{itra4b}). Set $\kappa=0$.
		\State \textbf{Repeat until convergence of the objective function in (\ref{itra4})/(\ref{sitra4}):}
Update $\gammak_k$ by (\ref{itra5}) for the GM-rate (\ref{itra4}) or $\gammak_k\equiv 1$ for the SR (\ref{sitra4}).
Generate the composite real form $\vkoa$ of $\hwkoa$ by (\ref{iaa13}) and the composite real form
$\vkoe$ of $\hwkoe$ by (\ref{tae13}). Reset $\kappa\leftarrow \kappa+1$.
 \State \textbf{Output} $\hwk\triangleq (\hwka,\hwke)$.
	\end{algorithmic}
\end{algorithm}
\section{Simulation Results}
The performance of the proposed algorithms along with their convergence is investigated by numerical examples in this section. The $8\times 8$-URA BS is deployed at the center of a cell  with a radius of 250 meters, where $K=30$ UEs are randomly placed. The height of BS antennas is $25$ meters, while the height of UEs is $1.5$ meters.

The channel $h_k$ spanning from the BS to UE $k$ is represented by the correlated Rayleigh fading model given as \cite{Adetal13,NKA19},
\begin{equation}
	h_k=\sqrt{10^{-\rho_k/10}} \mathrm{R}_k^{\frac{1}{2}}\tilde{h}_k\in\mathbb{C}^{M^2},
\end{equation}
where $\rho_k$ is the path-loss and shadow-fading coefficient, $\mathrm{R}_k\in\mathbb{C}^{M^2\times M^2}$ is the correlation matrix of channel $h_k$, and $\tilde{h}_k\in\mathcal{C}(0,I_{M^2})$. Following \cite{Yinetal14}, the correlation between the $(m,n)$-th antenna element and the $(p,q)$-th antenna element, with the $(m,n)$-th antenna element indicating the $m$-th in elevation and $n$-th in azimuth of the URA, is given as
\begin{equation}
	[\mathrm{R}_k]_{(m,n),(p,q)}=\frac{\gamma_1}{\sqrt{\gamma_5}}e^{-\frac{\gamma_7}{2\gamma_5}}e^{j\frac{\gamma_2\gamma_6}{\gamma_5}}e^{-\frac{(\gamma_2\sigma_{\alpha}\sin\alpha_k)^2}{2\gamma_5}},
\end{equation}
where
$\gamma_1=e^{j\pi(p-m)\cos\beta_k}e^{-\frac{1}{2}[\sigma_{\beta}\pi(p-m)\sin\beta_k]^2}$, $\gamma_2=\pi(q-n)\sin\beta_k$, $\gamma_3=\sigma_{\beta}\pi(q-n)\cos\beta_k$, $\gamma_4=\frac{1}{2}(\sigma_{\beta}\pi)^2(p-m)(q-n)\sin2\beta_k$, $\gamma_5=\gamma_3^2\sigma_{\alpha}^2\sin^2\alpha_k+1$, $\gamma_6=\gamma_4\sigma_{\alpha}^2\sin^2\alpha_k+\cos\alpha_k$, and $\gamma_7=\gamma_3^2\cos^2\alpha_k-\gamma_4^2\sigma_{\alpha}^2\sin^2\alpha_k-2\gamma_4\cos\alpha_k$. Furthermore, $\alpha_k$ and $\beta_k$ are the azimuth and elevation angles of UE $k$ geometrically determined based on the UE's location relative to the BS, respectively, while $\sigma_{\alpha}$ and $\sigma_{\beta}$ are the angular spreads in the azimuth and elevation domains set to $\sigma_{\alpha}=\sigma_{\beta}=5^{\circ}$, respectively.
The path-loss and shadow-fading of UE $k$ at a distance $d_k$ from the BS is set to $\rho_{k}=19.56+39.08\log_{10}(d_k)+\xi_k$ (in dB), where $\xi_k\sim \mathcal{N}(0,\sigma_\mathrm{sf}^2)$ is the shadow-fading coefficient with $\sigma_\mathrm{sf}=6$ \cite{3GPP}. The carrier frequency is set to $2$ GHz, the background noise power density is set to $-174$ dBm/Hz, and the bandwidth is set to $B = 10$ MHz. The tolerance introduced for declaring the algorithms' convergence is $1e-3$. The results are multiplied by $\log_2e$ to convert the unit of nats/sec into the unit of bps/Hz.

\begin{table*}[!t]
	\centering
	\caption{The average number of near-zero rate UEs at $P = 30$ dBm}
	\begin{tabular}{|c|c|c|c|c|c|}
		\hline
		& 2D-Q1 & 2D-Q2 & FD & IGS-2D-Q1 & IGS-2D-Q2 \\ \hline
		The average number of near-zero rate UEs & 15 & 13 & 13 & 12 & 7 \\ \hline
	\end{tabular}
	\label{table:num_zero_SR}
\end{table*}

\begin{table*}[!t]
	\centering
	\caption{The MR and SR achieved by the GM-rate and MR algorithms at $P = 30$ dBm}
	\begin{tabular}{|l|c|c|c|c|c|c|c|c|c|}
		\hline
		& GM-Q1   & GM-Q2   & GM-FD   & IGS-GM-Q1   & IGS-GM-Q2   & MR-Q1   & MR-Q2   & MR-FD   & IGS-MR-Q1    \\ \hline
		MR (bps/Hz) & 0.7198  & 0.8096  & 0.8987   & 1.1261    & 2.4174  & 1.6682  & 2.3139  & 2.8401  & 2.8471    \\ \hline
		SR (bps/Hz) & 85.2776 & 126.7677 & 200.8374 & 119.2307  & 154.0296 & 50.0701  & 69.4276 & 85.2564  & 85.4687 \\ \hline
	\end{tabular}
	\label{table:MR_SR_M8_P30}
\end{table*}

\begin{figure*}[!t]
	\centering
	\begin{minipage}[h]{0.48\textwidth}
		\centering
		\includegraphics[width=0.9 \textwidth]{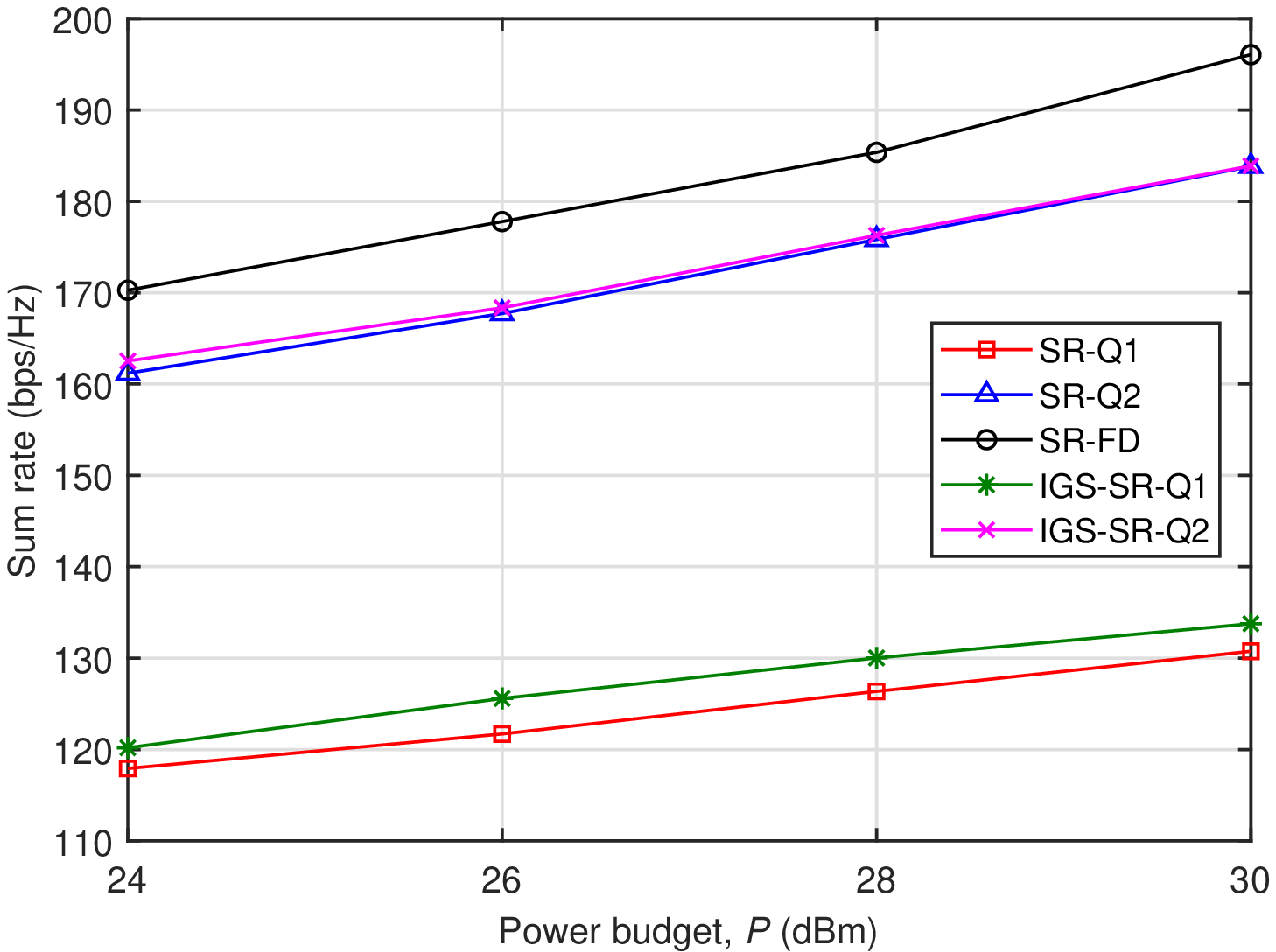}
		\caption{The SR vs. power budget $P$.}
		\label{fig:M8_sum_rate_vs_P_by_SR}
	\end{minipage}
	\hspace{0.3cm}
	\begin{minipage}[h]{0.48\textwidth}
		\centering
		\includegraphics[width=0.9 \textwidth]{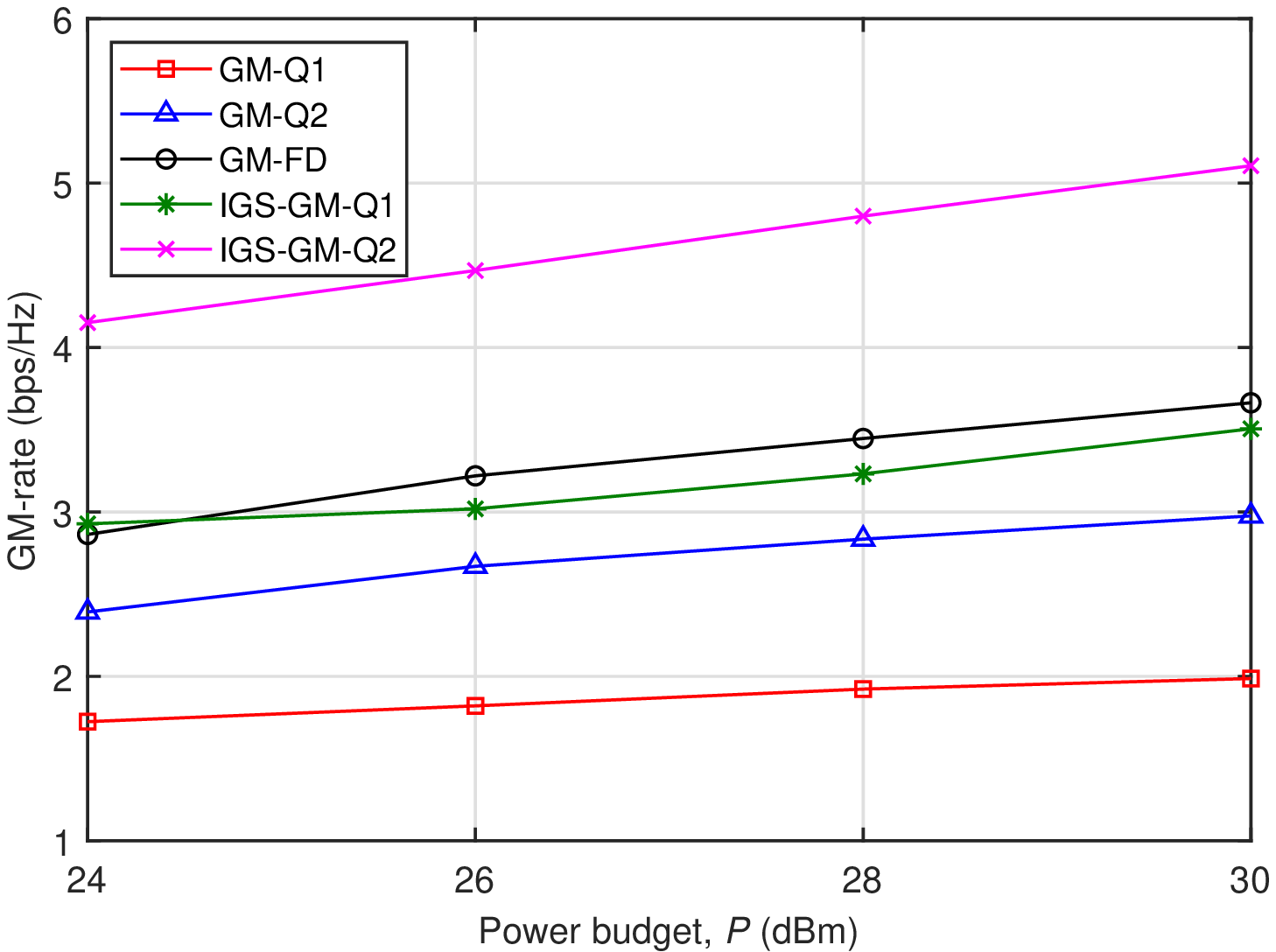}
		\caption{The GM-rate vs. power budget $P$.}
		\label{fig:M8_GM_rate_vs_P}
	\end{minipage}
	\vspace{-0.3cm}
\end{figure*}

\begin{figure*}[!t]
	\centering
	\begin{minipage}[h]{0.48\textwidth}
		\centering
		\includegraphics[width=0.9 \textwidth]{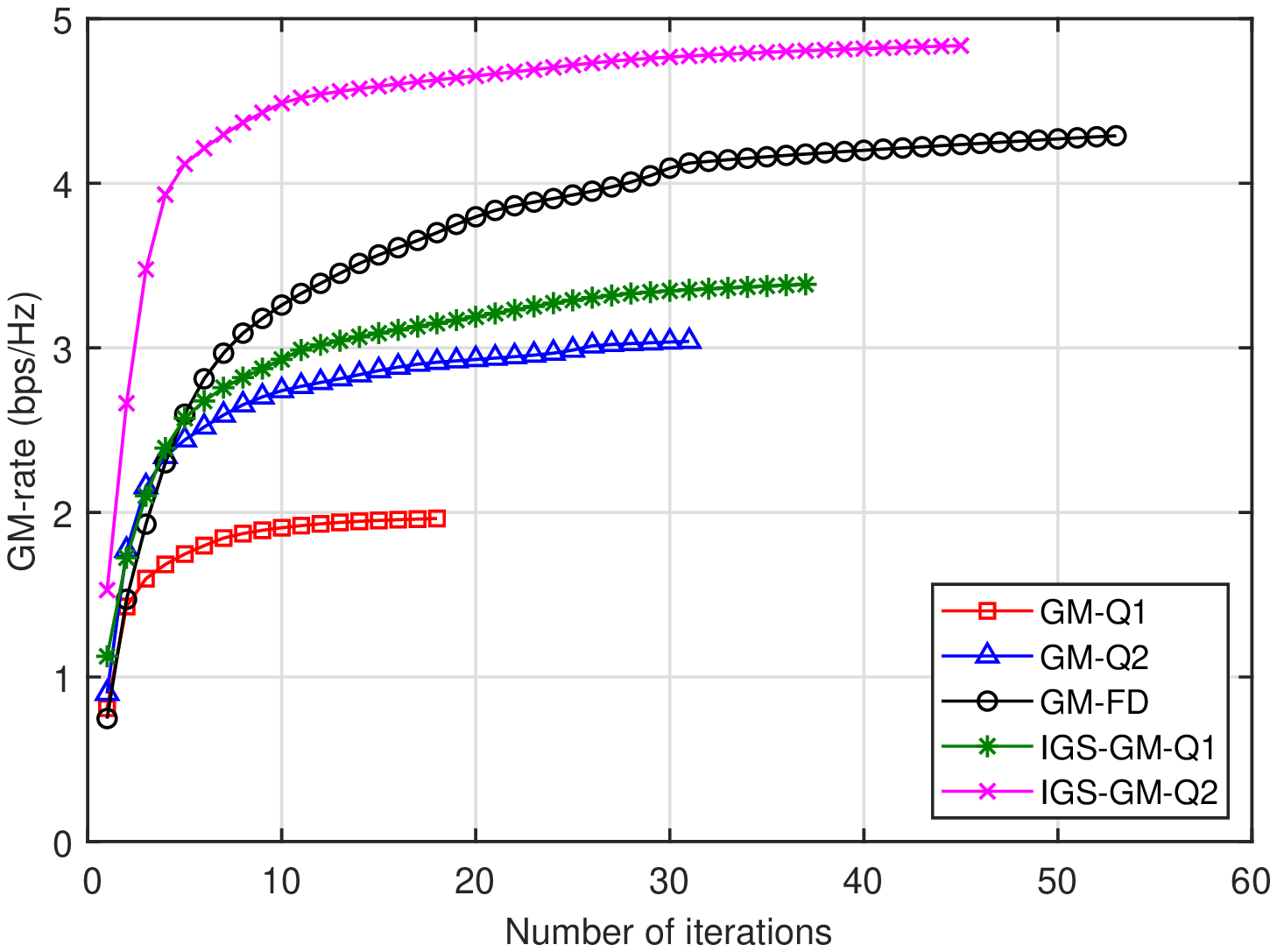}
		\caption{The GM-rate convergence of the GM-rate algorithms at $P = 30$ dBm.}
		\label{fig:M8_conv_GM_rate}
	\end{minipage}
	\hspace{0.3cm}
	\begin{minipage}[h]{0.48\textwidth}
		\centering
		\includegraphics[width=0.9 \textwidth]{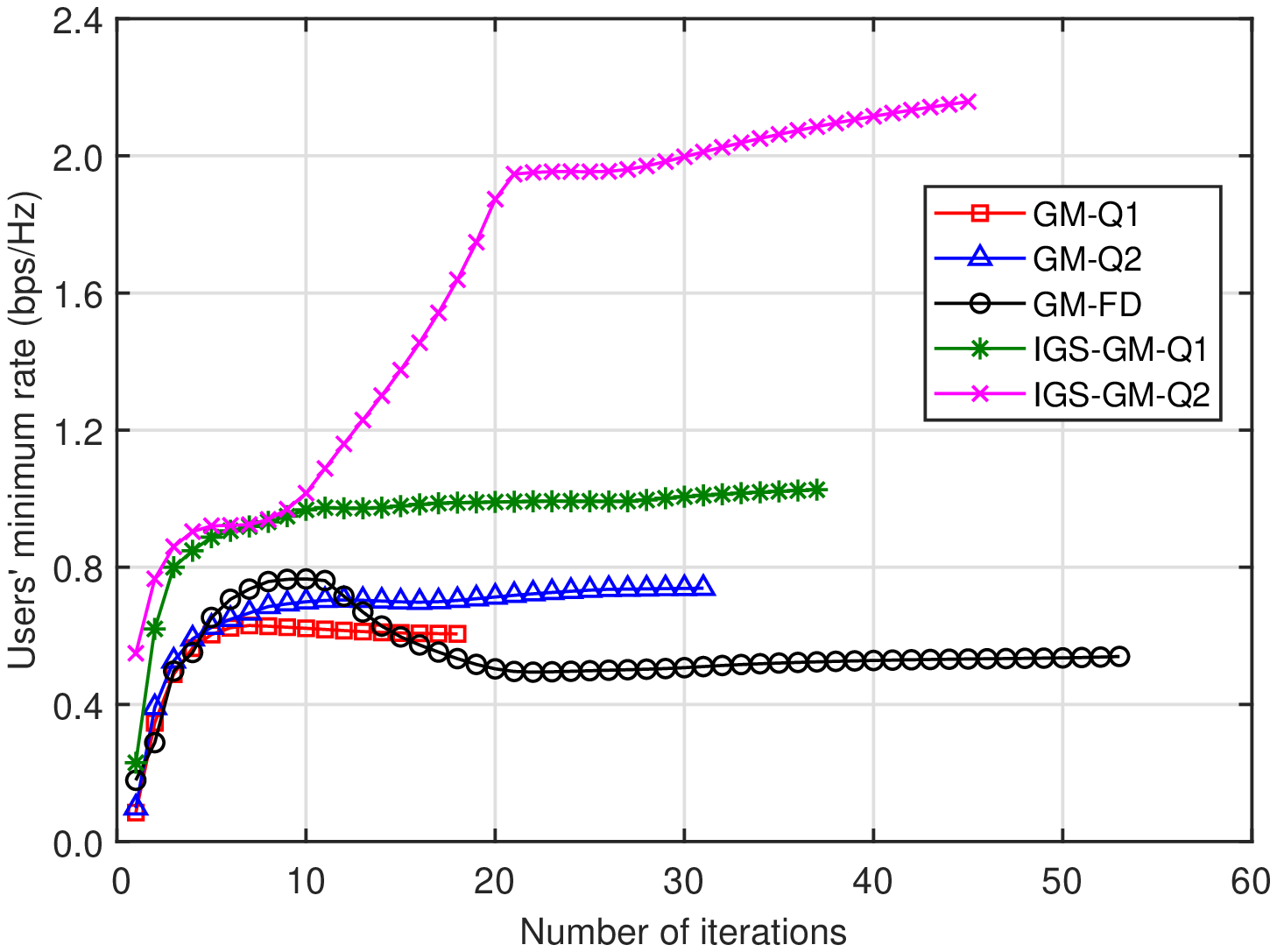}
		\caption{The MR convergence of the GM-rate algorithms at $P = 30$ dBm.}
		\label{fig:M8_conv_min_rate_by_GM}
	\end{minipage}
	\vspace{-0.3cm}
\end{figure*}

\begin{figure*}[!t]
	\centering
	\begin{minipage}[h]{0.48\textwidth}
		\centering
		\includegraphics[width=0.9 \textwidth]{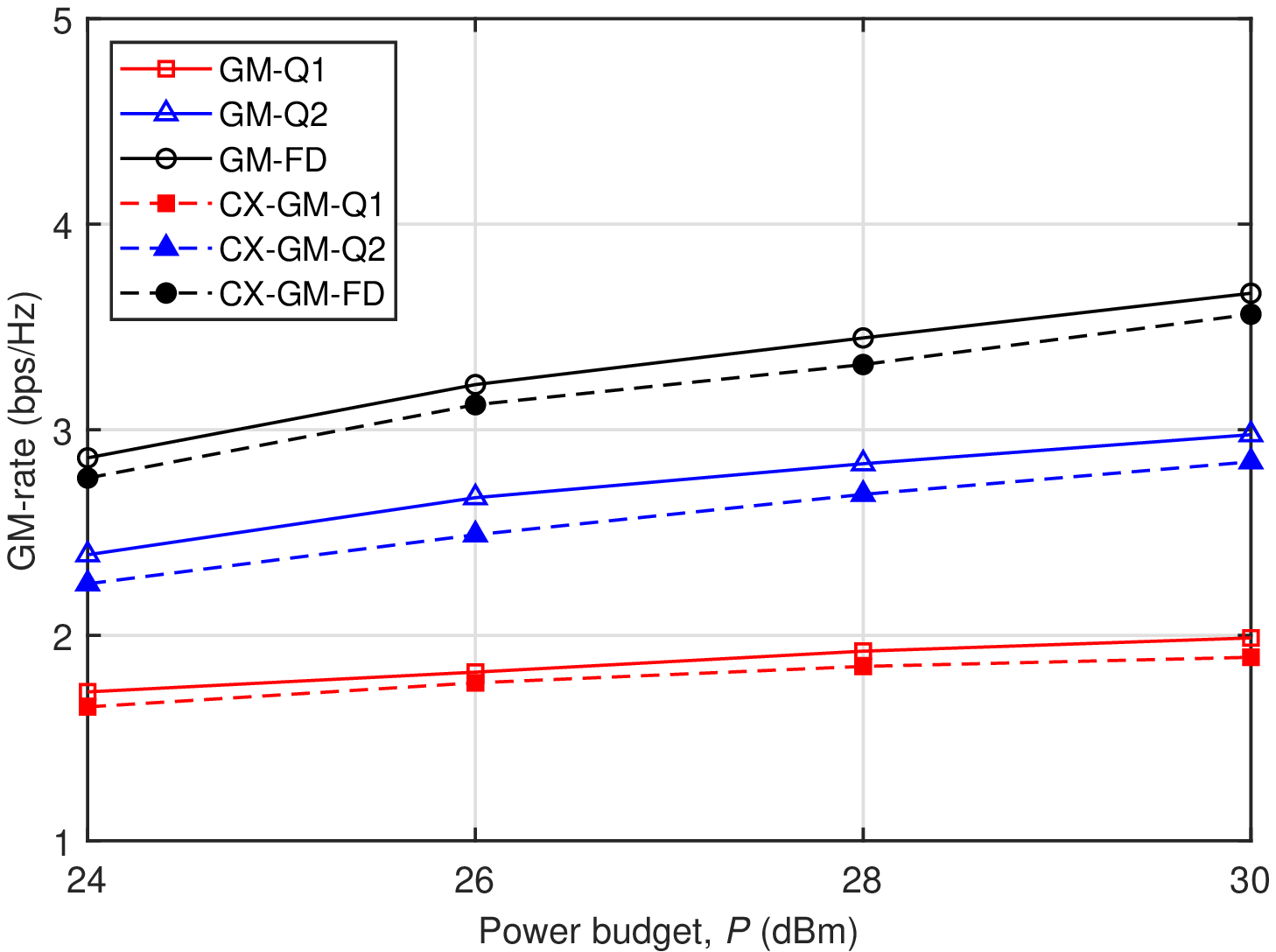}
		\caption{The GM-rate vs. power budget $P$ of the closed-form and convex-solver based algorithms.}
		\label{fig:M8_GM_rate_vs_P_by_closedform_cvx}
	\end{minipage}
	\hspace{0.3cm}
	\begin{minipage}[h]{0.48\textwidth}
		\centering
		\includegraphics[width=0.9 \textwidth]{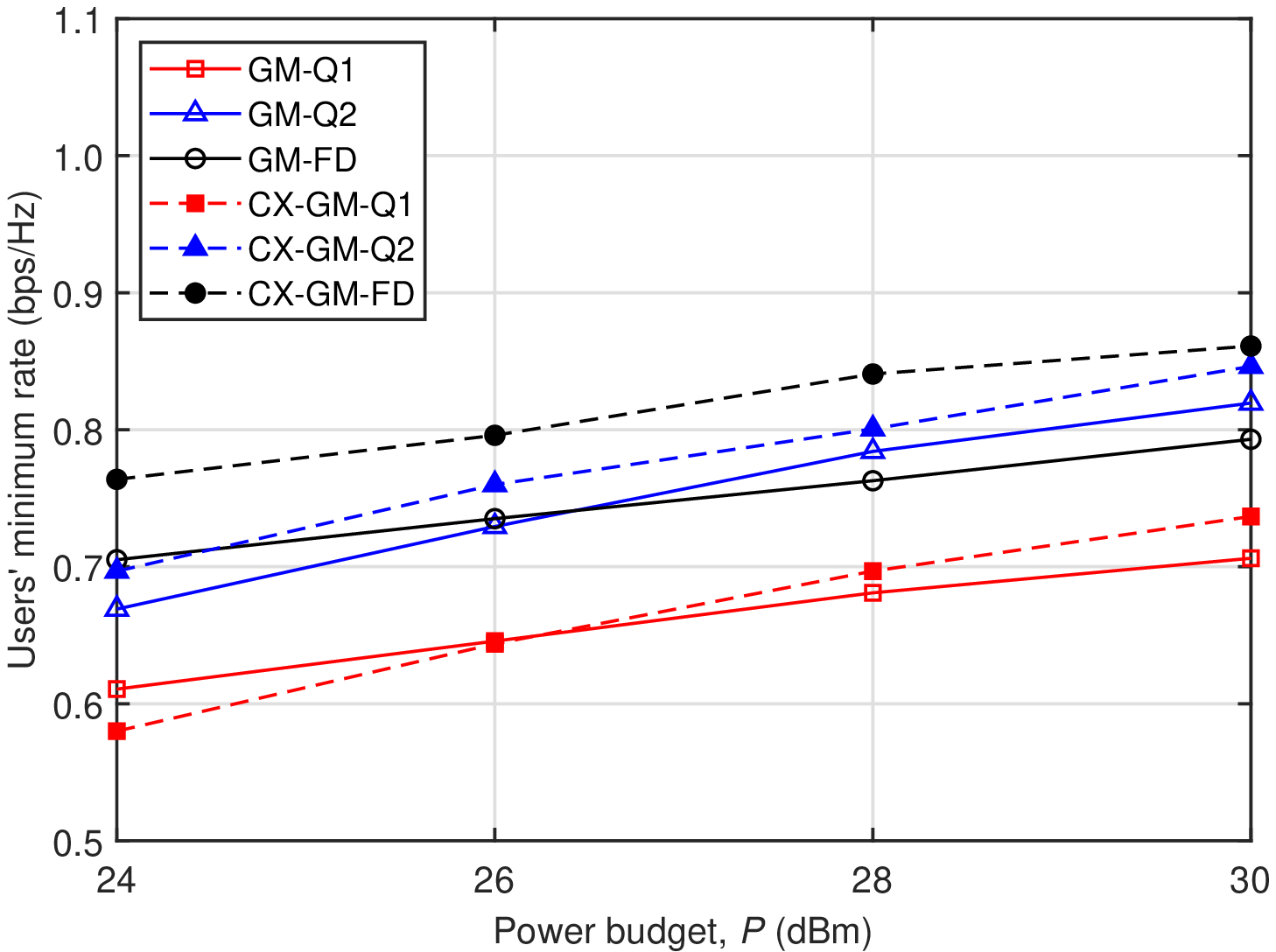}
		\caption{The MR vs. power budget $P$ of the closed-form and convex-solver based algorithms.}
		\label{fig:M8_min_rate_vs_P_by_closedform_cvx}
	\end{minipage}
	\vspace{-0.3cm}
\end{figure*}

\begin{figure*}[!t]
	\centering
	\begin{minipage}[h]{0.48\textwidth}
		\centering
		\includegraphics[width=0.9 \textwidth]{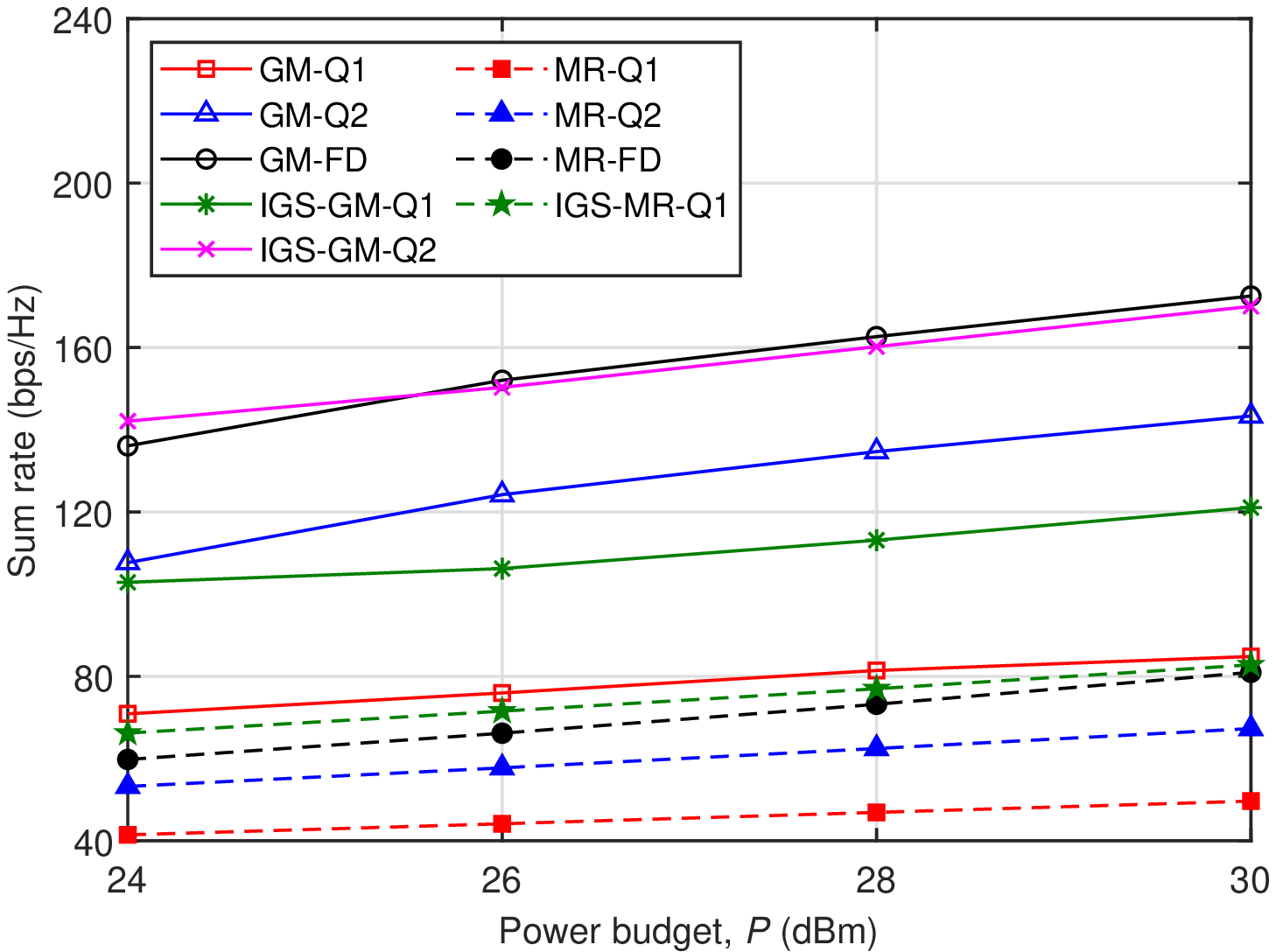}
		\caption{The SR vs. power budget $P$ by the GM-rate and MR algorithms.}
		\label{fig:M8_sum_rate_vs_P_by_GM_and_MR}
	\end{minipage}
	\hspace{0.3cm}
	\begin{minipage}[h]{0.48\textwidth}
		\centering
		\includegraphics[width=0.9 \textwidth]{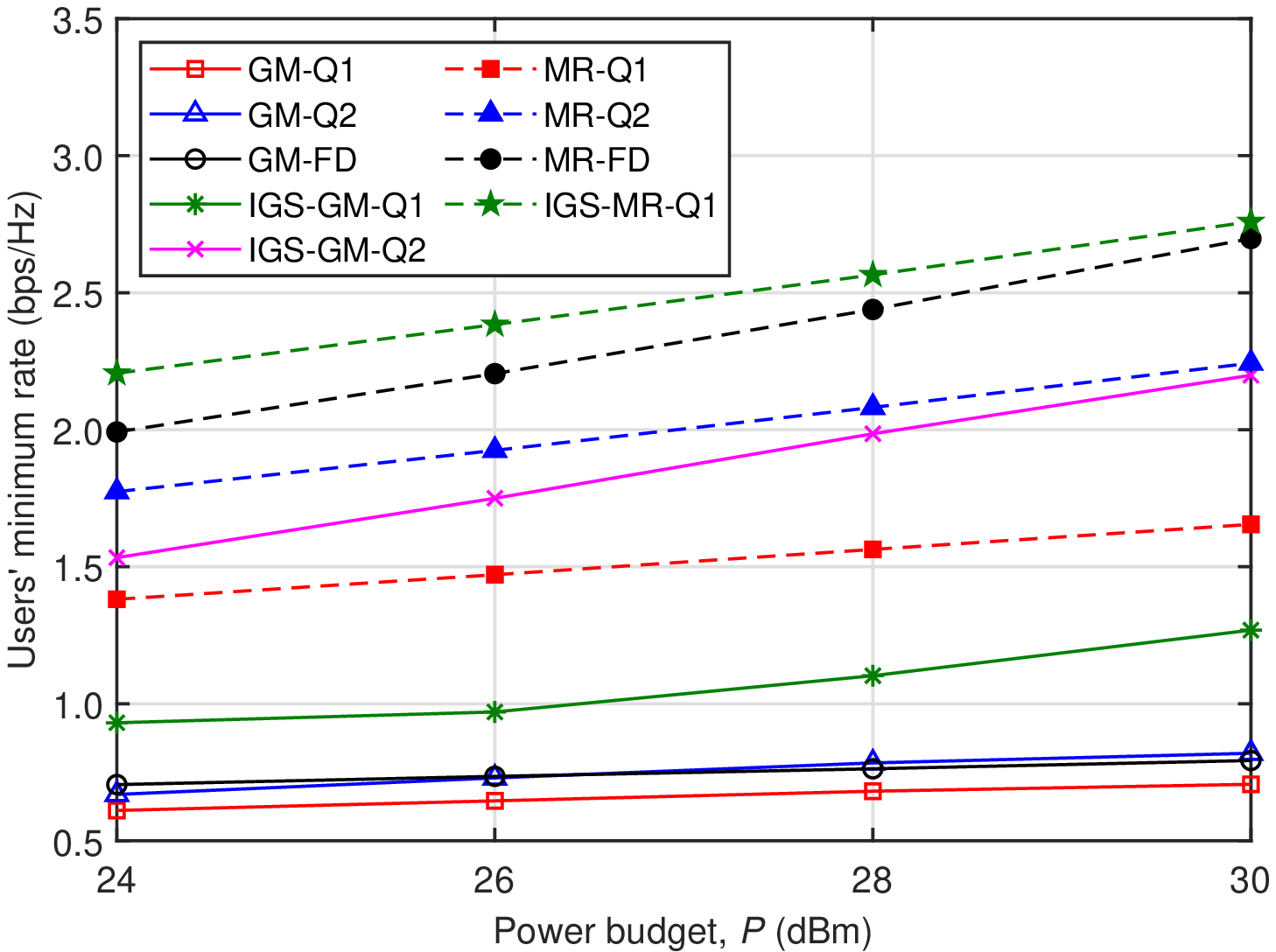}
		\caption{The MR vs. power budget $P$ by the GM-rate and MR algorithms.}
		\label{fig:M8_min_rate_vs_P_by_GM_and_MR}
	\end{minipage}
	\vspace{-0.3cm}
\end{figure*}

\begin{figure*}[!t]
	\centering
	\begin{minipage}[h]{0.48\textwidth}
		\centering
		\includegraphics[width=0.9 \textwidth]{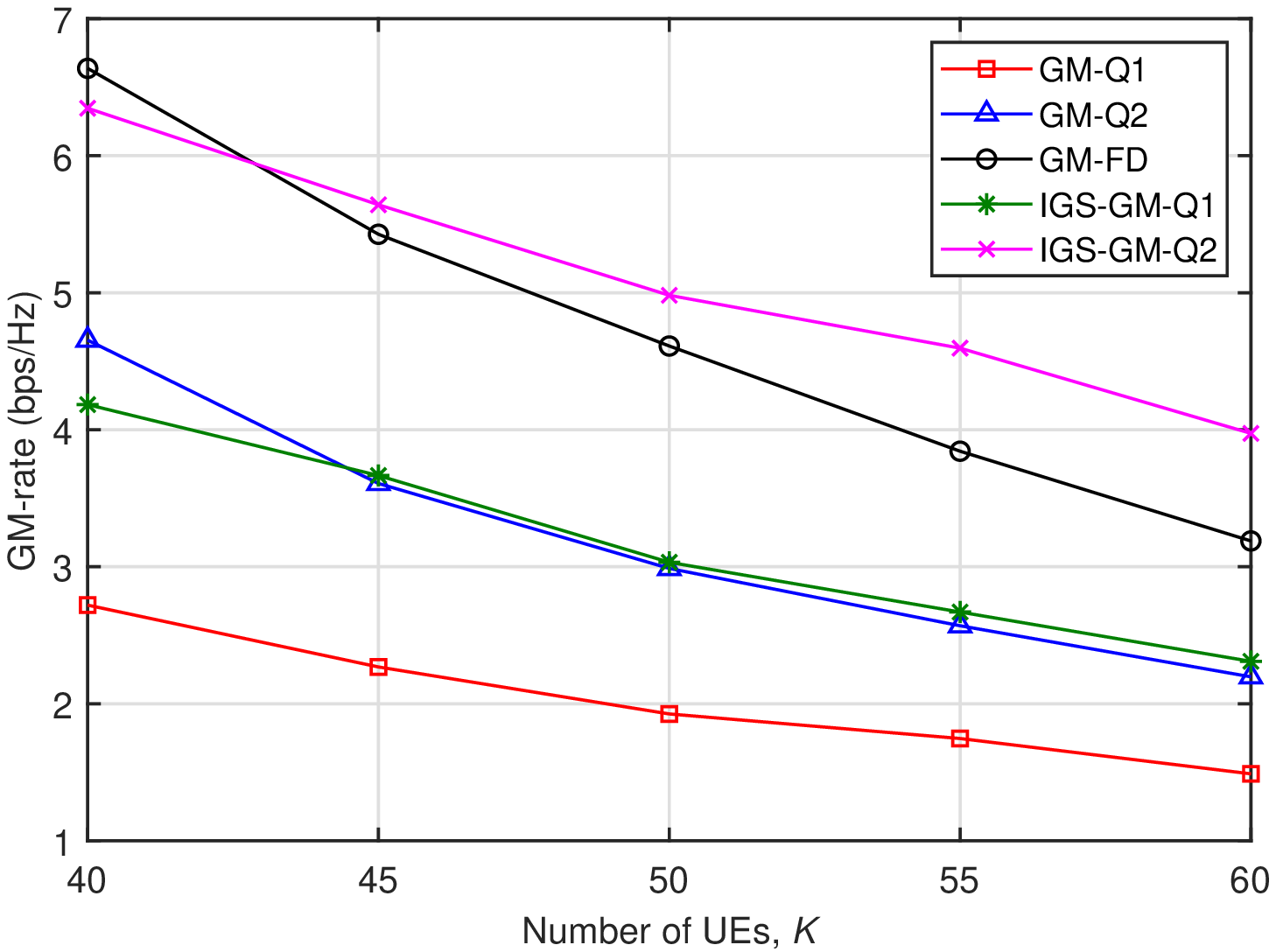}
		\caption{The GM-rate vs. the number $K$ of UEs.}
		\label{fig:M12_GM_rate_vs_UE}
	\end{minipage}
	\hspace{0.3cm}
	\begin{minipage}[h]{0.48\textwidth}
		\centering
		\includegraphics[width=0.9 \textwidth]{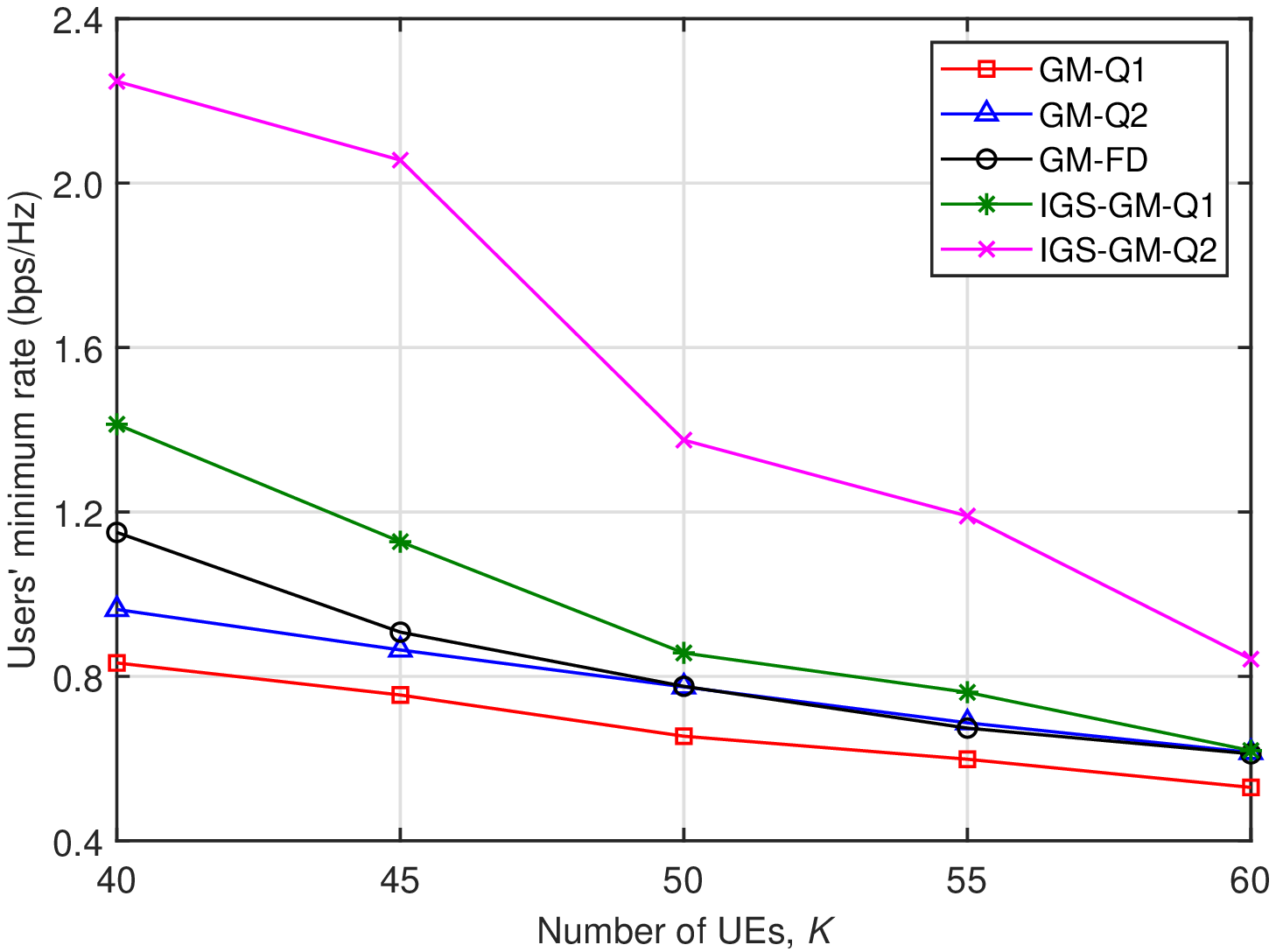}
		\caption{The MR vs. the number $K$ of UEs.}
		\label{fig:M12_min_rate_vs_UE}
	\end{minipage}
	\vspace{-0.3cm}
\end{figure*}

We use the following legends to specify the proposed implementation:
MR-Q1/Q2 refer to the convex-solver based  Alg. \ref{alg4m} with $Q=1$ (2D beamforming)/$Q=2$, while MR-FD refers to the baseline convex-solver based Alg. \ref{alg7m};
SR-Q1/GM-Q1 and SR-Q2/GM-Q2 refer to the closed-form based Alg. \ref{alg4} for SR/GM-rate maximization associated with $Q=1$ and $Q=2$, while SR/GM-FD refers to the baseline closed-form based Alg. \ref{alg7} for SR/GM maximization; CX-GM-Q1/Q2 refers to the
convex-solver based Alg. \ref{alg4gm} for GM-rate maximization with $Q=1/Q=2$;
IGS-MR-Q1 refers to the convex-solver based Alg.  \ref{mialg4} for $Q=1$.\footnote{Due to the high computational complexity of (\ref{comc}), Alg.  \ref{mialg4} 
becomes excessively complex for the large-scale simulations of this section.}
Lastly, IGS-SR/GM-Q1/Q2 refer to the closed-form based IGS Alg. \ref{ialg4} for SR and GM-rate maximization associated with $Q=1/Q=2$.

We start by characterizing the SR maximization. Fig. \ref{fig:M8_sum_rate_vs_P_by_SR} shows the
SR vs. power budget $P$. One can see that the SR-FD achieves the highest SR, followed by SR-Q2 and IGS-SR-Q2. However, Table \ref{table:num_zero_SR} shows that SR maximization results in many near-zero rates, and thus effectively disconnects many users. Therefore, the SR maximization is not applicable in multi-user scenarios. Thus, from now we focus our attention on simulating the MR and GM-rate maximization only with the following interesting outcomes:
$(i)$ GM-rate maximization indeed achieves both good MR and SR, i.e. it provides an Pareto optimal solution for optimizing multi-objective MR and SR, while
MR maximization sacrifices the SR to support the MR;
$(ii)$ The closed-form based GM-Q1/Q2 performs similarly well to the convex-solver based CX-GM-Q1/Q2, but the computational complexity of the former is linear;
$(iii)$ GM-Q2 already approaches GM-FD, and
$(iv)$ IGS-GM-Q2 outperforms GM-FD.

Fig. \ref{fig:M8_GM_rate_vs_P} plots the GM-rate vs. power budget $P$. IGS-GM-Q1 gradually approaches the baseline GM-FD, while IGS-GM-Q2 significantly outperforms GM-FD.   Fig.  \ref{fig:M8_conv_GM_rate} shows the GM-rate convergence at $P = 30$ dBm. GM-FD requires almost 50 iterations for convergence, because it has the largest numbers of decision variables. By contrast, GM-Q1 has the smallest number of decision variables, and it requires less than 20 iterations to achieve convergence. 
	Fig.  \ref{fig:M8_conv_min_rate_by_GM} shows the MR behaviour of the closed-form expression based GM-rate algorithms at $P = 30$ dBm, where GM-Q1/Q2 tend to slightly sacrifice the MR for the sake of achieving a higher GM-rate, while GM-FD substantially reduces the MR to allow the GM-rate to be increased further. By contrast, IGS-GM-Q1/Q2 improve the MR along with the GM-rate. GM-FD and IGS-GM-Q2 require more iterations for convergence than the others, because the former has a large number of decision variables.

We compare the performances of the closed-form expression based and convex-solver based algorithms.
Fig. \ref{fig:M8_GM_rate_vs_P_by_closedform_cvx} and Fig. \ref{fig:M8_min_rate_vs_P_by_closedform_cvx} plot the GM-rate and MR achieved, which show that they perform similarly although the computational complexity of the former is very high compared to that of the latter.

Furthermore, we compare the SR and MR achieved by MR and GM-rate maximization.  Fig. \ref{fig:M8_sum_rate_vs_P_by_GM_and_MR} shows that
the GM-rate maximization achieves higher SR than  MR maximization. Furthermore,
GM-FD and IGS-GM-Q2 achieve the highest SR, followed by GM-Q2 and IGS-GM-Q1.
Also  IGS-MR-Q1 achives a better SR than MR-Q2. 
Fig. \ref{fig:M8_min_rate_vs_P_by_GM_and_MR}, which plots the MRs achieved, shows that
MR maximization indeed achieves better MR than the GM-rate maximization, while
IGS-GM-Q2 barely catches up with GM-Q2 at $P=30$ dBm.  It is not surprising that
	IGS-MR-Q1 achieves the best MR and it even performs better than FD-MR does.
Fig. \ref{fig:M8_conv_GM_rate}, Fig. \ref{fig:M8_conv_min_rate_by_GM}, Fig. \ref{fig:M8_sum_rate_vs_P_by_GM_and_MR}, and Fig. \ref{fig:M8_min_rate_vs_P_by_GM_and_MR} reveal that by iterating upon evaluating the GM-rate maximization, one can satisfy the  MR required, while maintaining a good SR. Hence, for obtaining a reasonable MR at a moderate computational overhead, we prefer the closed-form expression based GM-rate algorithms, instead of the convex-solver based MR algorithms.

Table \ref{table:MR_SR_M8_P30} provides the MR and SR achieved by the proposed algorithms at $P = 30$ dBm for a particular channel generation, which confirms that the GM-rate maximization associated with linear computational complexity  achieves similar MRs to that of direct
MR maximization, but the latter imposes high-order polynomially increasing complexity.
While the former succeeds in maintaining a very good SR, the latter has to sacrifice the SR in favor of maximizing the MR.

For more qualitative analysis, Table \ref{table:rate_M8} shows the min-rate/max-rate patterns and Jain's fairness index of the user rate, which is defined as $\frac{(\sum_{k=1}^{K}r_k)^2}{K\sum_{k=1}^{K}r_k^2}$ \cite{Jain84}, at $P = 30$ dBm. The IGS beamformers achieve higher min-rate to max-rate ratio and higher Jain's fairness index, demonstrating that they promise fairer rate distributions among the UEs than others.

To substantiate the superior capability of GM-rate maximization in terms of efficient beamforming at identical transmit powers at the antennas, we provide Table \ref{table:power_M8} for the min-power to max-power ratio and for Jain's fairness index of the antenna transmit power distributions. Observe in Table \ref{table:power_M8} that the antenna transmit power distributions achieved by the sum of one or two outer product-based beamformers are similar to that achieved by FD.

\begin{table*}[!t]
	\centering
	\caption{Min-rate/max-rate and Jain's fairness index of user rate by the GM-rate algorithms at $P = 30$ dBm and $K = 30$}
	\begin{tabular}{|l|c|c|c|c|c|}
		\hline
		& GM-Q1 & GM-Q2 & GM-FD & IGS-GM-Q1  & IGS-GM-Q2 \\ \hline
		Min-rate/max-rate          & 0.0547		& 0.0486	& 0.0459  & 0.1395  & 0.1532 \\ \hline
		Jain's fairness index      & 0.4725		& 0.4944	& 0.5561  & 0.7788  & 0.7923 \\ \hline
	\end{tabular}
	\label{table:rate_M8}
\end{table*}

\begin{table*}[!t]
	\centering
	\caption{Min-power/max-power and Jain's fairness index of antenna power by the GM-rate algorithms at $P = 30$ dBm and $K = 30$}
	\begin{tabular}{|l|c|c|c|c|c|}
		\hline
		& GM-Q1 & GM-Q2 & GM-FD & IGS-GM-Q1  & IGS-GM-Q2 \\ \hline
		Min-power/max-power        & 0.1922		& 0.2008	& 0.1873  & 0.1512  & 0.1865 \\ \hline
		Jain's fairness index      & 0.8195		& 0.8148	& 0.7949  & 0.7946  & 0.7985 \\ \hline
	\end{tabular}
	\label{table:power_M8}
\end{table*}

\begin{table*}[!t]
	\centering
	\caption{Min-rate/max-rate and Jain's fairness index of user rate by the GM-rate algorithms at $P = 40$ dBm and $K = 60$}
	\begin{tabular}{|l|c|c|c|c|c|}
		\hline
		& GM-Q1 & GM-Q2 & GM-FD & IGS-GM-Q1  & IGS-GM-Q2 \\ \hline
		Min-rate/max-rate          & 0.0313		& 0.0290	& 0.0291  & 0.0520  & 0.0506 \\ \hline
		Jain's fairness index      & 0.3577		& 0.3768	& 0.4875  & 0.5862  & 0.6992 \\ \hline
	\end{tabular}
	\label{table:rate_M12}
\end{table*}

\begin{table*}[!t]
	\centering
	\caption{Min-power/max-power and Jain's fairness index of antenna power by the GM-rate algorithms at $P = 40$ dBm and $K = 60$}
	\begin{tabular}{|l|c|c|c|c|c|}
		\hline
		& GM-Q1 & GM-Q2 & GM-FD & IGS-GM-Q1  & IGS-GM-Q2 \\ \hline
		Min-power/max-power        & 0.1717		& 0.1883	& 0.2051  & 0.1332  & 0.1915 \\ \hline
		Jain's fairness index      & 0.8019		& 0.8165	& 0.8085  & 0.7989  & 0.8300 \\ \hline
	\end{tabular}
	\label{table:power_M12}
\end{table*}

We now further evaluate the performance of the GM-rate algorithms for larger numbers of UEs in the scenario of a $12 \times 12$ URA at the BS. In the following simulations, more UEs are randomly placed in a cell with a radius of 500 meters. The transmit power budget  is fixed at 40 dBm.

Fig. \ref{fig:M12_GM_rate_vs_UE} plots the GM-rate vs. the number $K$ of UEs. IGS-GM-Q2 achieves the highest GM-rate when serving more UEs. Fig. \ref{fig:M12_min_rate_vs_UE}, which plots the MR vs. the number $K$ of UEs, shows that IGS-GM-Q2 is also capable of reaching significantly higher MR than the others. Furthermore, IGS-GM-Q1 outperforms FD at smaller number of UEs. When the number of UEs increases to $K=60$, the MR achieved by GM-FD, IGS-GM-Q1, and GM-Q2 are getting close to each other.

The min-rate/max-rate patterns and Jain's fairness index of the user rate at $K = 60$ are also included in Table \ref{table:rate_M12}. We can see that the IGS-based algorithms are capable of reaching fairer user rate distributions even when serving a large number of UEs.

According to the min-power to max-power ratio and to Jain's fairness index provided in Table \ref{table:power_M12}, we observe that the antenna transmit power distributions achieved by the sum of one or two outer product-based beamformers are similar to that attained by FD. Table \ref{table:power_M12} follows
a similar trend  to Table \ref{table:power_M8} provided for an $8\times 8$ URA.

\section{Conclusions}
A suite of FD m-MIMO systems was conceived for a base station employing $M\times M$-URAs in the downlink to support multiple users.  We proposed a low-complexity class of beamformers,
which are represented by the sums of outer products of $M$-dimensional azimuth and $M$-dimensional elevation beamformers.
We have developed low-complexity algorithms for maximizing the SR, MR and GM-rate and  showed that the
sums of two outer products associated with the design complexity order of $M$ perform similarly to the baseline FD beamformer having
the excessive design complexity order  of $M^2$. Furthermore, the scheme relying on the sum of two outer products in improper Gaussian signaling succeeds in outperforming the baseline FD beamformer.

\bibliographystyle{ieeetr}
\balance \bibliography{mmwave}

\begin{thebibliography}{10}

\bibitem{Nametal13}
{Y.-H. Nam et al.}, ``Full-dimension {MIMO (FD-MIMO)} for next generation
  cellular technology,'' {\em IEEE Commun. Mag.}, vol.~51, pp.~172--179, Jun.
  2013.

\bibitem{Kimetal14}
{Y. Kim et al.}, ``Full dimension mimo {(FD-MIMO)}: the next evolution of
  {MIMO} in {LTE} systems,'' {\em IEEE Wirel. Commun.}, vol.~21, pp.~26--33,
  Apr. 2014.

\bibitem{Jietal17}
{H. Ji et al.}, ``Overview of full-dimension {MIMO} in {LTE}-advanced pro,''
  {\em IEEE Commun. Mag.}, vol.~55, pp.~176--184, Feb 2017.

\bibitem{Monetal15}
{B. Mondal et al.}, ``{3D channel model in 3GPP},'' {\em IEEE Commun. Mag.},
  vol.~53, pp.~16--23, Mar. 2015.

\bibitem{Lietal16tvt}
X.~Li, S.~Jin, X.~Gao, and R.~W. Heath, ``Three-dimensional beamforming for
  large-scale {FD-MIMO} systems exploiting statistical channel state
  information,'' {\em IEEE Trans. Vehic. Tech.}, vol.~65, pp.~8992--9005, Nov.
  2016.

\bibitem{Liuetal17}
W.~Liu, Z.~Wang, C.~Sun, S.~Chen, and L.~Hanzo, ``Structured non-uniformly
  spaced rectangular antenna array design for {FD-MIMO} systems,'' {\em IEEE
  Trans. Wirel. Commun.}, vol.~16, pp.~3252--3265, May 2017.

\bibitem{NKDA18}
Q.-U.-A. Nadeem, A.~Kammoun, M.~Debbah, and M.-S. Alouini, ``Design of {5G}
  full dimension massive {MIMO} systems,'' {\em IEEE Trans. Commun.}, vol.~66,
  pp.~726--740, Feb. 2018.

\bibitem{NKA19}
Q.-U.-A. Nadeem, A.~Kammoun, and M.-S. Alouini, ``Elevation beamforming with
  full dimension {MIMO} architectures in {5G} systems: A tutorial,'' {\em IEEE
  Commun. Surv. Tut.}, vol.~21, no.~4, pp.~3238--3273, 2019.

\bibitem{LLQJ20}
X.~Li, Z.~Liu, N.~Qin, and S.~Jin, ``{FFR} based joint {3D} beamforming
  interference coordination for multi-cell {FD-MIMO} downlink transmission
  systems",'' {\em IEEE Trans. Vehic. Techn.}, vol.~69, pp.~3105--3118, Mar.
  2020.

\bibitem{NTDP19}
L.~D. Nguyen, H.~D. Tuan, T.~Q. Duong, and H.~V. Poor, ``Multi-user regularized
  zero forcing beamforming,'' {\em IEEE Trans. Signal Process.}, vol.~67,
  pp.~2839--2853, Jun. 2019.

\bibitem{Yuetaltvt20}
H.~Yu, H.~D. Tuan, A.~A. Nasir, T.~Q. Duong, and L.~Hanzo, ``Improper
  {G}aussian signaling for computationally tractable energy and information
  beamforming,'' {\em IEEE Trans. Veh. Techn.}, vol.~69, pp.~13990--13995, Nov.
  2020.

\bibitem{Nguetal21}
L.~D. Nguyen, H.~D. Tuan, T.~Q. Duong, H.~V. Poor, and L.~Hanzo,
  ``Energy-efficient multi-cell massive {MIMO} subject to minimum user-rate
  constraints,'' {\em IEEE Trans. Commun.}, vol.~69, pp.~914--928, Feb. 2021.

\bibitem{Yinetal14}
D.~Ying, F.~W. Vook, T.~A. Thomas, D.~J. Love, and A.~Ghosh, ``Kronecker
  product correlation model and limited feedback codebook design in a {3D}
  channel model,'' in {\em Proc. 2014 IEEE Inter. Conf. Commun. (ICC)},
  pp.~5865--5870, 2014.

\bibitem{ALH17}
A.~Alkhateeb, G.~Leus, and R.~W. Heath, ``Multi-layer precoding: A potential
  solution for full-dimensional massive {MIMO} systems,'' {\em IEEE Trans.
  Wirel. Commun.}, vol.~16, pp.~5810--5824, Mar. 2017.

\bibitem{Kanetal17tvt}
J.~Kang, O.~Simeone, J.~Kang, and S.~Shamai, ``Layered downlink precoding for
  {C-RAN} systems with full dimensional {MIMO},'' {\em IEEE Trans. Vehic.
  Techn.}, vol.~66, pp.~2170--2182, Mar. 2017.

\bibitem{Wanetal17}
Z.~Wang, W.~Liu, C.~Qian, S.~Chen, and L.~Hanzo, ``Two-dimensional precoding
  for {3-D} massive {MIMO},'' {\em IEEE Trans. Vehic. Techn.}, vol.~66,
  pp.~5488--5493, June 2017.

\bibitem{Sonetal19}
Y.~Song, C.~Liu, W.~Wang, N.~Cheng, M.~Wang, W.~Zhuang, and X.~Shen, ``Domain
  selective precoding in {3-D} massive {MIMO} systems,'' {\em IEEE J. Select.
  Topics Signal Process.}, vol.~13, pp.~1103--1117, May 2019.

\bibitem{TTN16}
H.~H.~M. Tam, H.~D. Tuan, and D.~T. Ngo, ``Successive convex quadratic
  programming for quality-of-service management in full-duplex {MU-MIMO}
  multicell networks,'' {\em IEEE Trans. Commun.}, vol.~64, pp.~2340--2353,
  June 2016.

\bibitem{Naetal17tsp}
A.~A. Nasir, H.~D. Tuan, T.~Q. Duong, and H.~V. Poor, ``Secrecy rate
  beamforming for multicell networks with information and energy harvesting,''
  {\em IEEE Trans. Signal Process.}, vol.~65, no.~3, pp.~677--689, 2017.

\bibitem{Yuetaljsac20}
H.~Yu, H.~D. Tuan, A.~A. Nasir, T.~Q. Duong, and H.~V. Poor, ``Joint design of
  reconfigurable intelligent surfaces and transmit beamforming under proper and
  improper {G}aussian signaling,'' {\em IEEE J. Sel. Areas Commun.}, vol.~38,
  pp.~2589--2603, Nov. 2020.

\bibitem{Yuetaltwc22}
H.~Yu, H.~D. Tuan, E.~Dutkiewicz, H.~V. Poor, and L.~Hanzo, ``Maximizing the
  geometric mean of user-rates to improve rate-fairness: {P}roper vs. improper
  {G}aussian signaling,'' {\em IEEE Trans. Wirel. Commun.}, vol.~21, no.~1,
  pp.~295--309, 2022.

\bibitem{Zhuetal22tvt}
W.~Zhu, H.~D. Tuan, E.~Dutkiewicz, and L.~Hanzo, ``Collaborative beamforming
  aided fog radio access networks,'' {\em IEEE Trans. Veh. Techn.}, vol.~71,
  pp.~7805--7820, Jul. 2022.

\bibitem{Nasetal22tvt}
A.~A. Nasir, H.~D. Tuan, E.~Dutkiewicz, and L.~Hanzo, ``Finite-resolution
  digital beamforming for multi-user millimeter-wave networks,'' {\em IEEE
  Trans. Veh. Techn.}, vol.~71, Sept. 2022.

\bibitem{Nasetal22tcom}
A.~A. Nasir, H.~D. Tuan, E.~Dutkiewicz, H.~V. Poor, and L.~Hanzo,
  ``Low-resolution {RIS}-aided multi-user {MIMO} signaling,'' {\em IEEE Trans.
  Commun.}, vol.~70, pp.~6517--6531, Oct. 2022.

\bibitem{TT13}
H.~Tuy and H.~D. Tuan, ``Generalized {S}-lemma and strong duality in nonconvex
  quadratic programming,'' {\em J. of Global Optimization}, vol.~56,
  pp.~1045--1072, 2013.

\bibitem{Tuybook}
H.~Tuy, {\em Convex Analysis and Global Optimization (second edition)}.
\newblock Springer International, 2017.

\bibitem{HJU13}
C.~Hellings, M.~Joham, and W.~Utschick, ``{QoS} feasibility in {MIMO} broadcast
  channels with widely linear transceivers,'' {\em IEEE Signal Process. Lett.},
  vol.~20, pp.~1134--1137, Nov. 2013.

\bibitem{Zeetal13}
Y.~Zeng, C.~M. Yetis, E.~Gunawan, Y.~L. Guan, and R.~Zhang, ``Transmit
  optimization with improper {G}aussian signaling for interference channels,''
  {\em IEEE Trans. Signal Process.}, vol.~61, pp.~2899--2913, Jun. 2013.

\bibitem{LAV16}
S.~Lagen, A.~Agustin, and J.~Vidal, ``On the superiority of improper {G}aussian
  signaling in wireless interference {MIMO} scenarios,'' {\em IEEE Trans.
  Commun.}, vol.~64, pp.~3350--3368, Aug. 2016.

\bibitem{NTDP19spl}
A.~A. Nasir, H.~D. Tuan, T.~Q. Duong, and H.~V. Poor, ``Improper {G}aussian
  signaling for broadcast interference networks,'' {\em IEEE Signal Process.
  Lett.}, vol.~26, pp.~808--812, Jun. 2019.

\bibitem{Tuetal19}
H.~D. Tuan, A.~A. Nasir, H.~H. Nguyen, T.~Q. Duong, and H.~V. Poor,
  ``Non-orthogonal multiple access with improper {G}aussian signaling,'' {\em
  IEEE J. Selec. Topics Signal Process.}, vol.~13, pp.~496--507, Mar. 2019.

\bibitem{RTN12}
U.~Rashid, H.~D. Tuan, and H.~H. Nguyen, ``Relay beamforming designs in
  multi-user wireless relay networks based on throughput maximin
  optimization,'' {\em IEEE Trans. Commun.}, vol.~61, pp.~1739--1749, May 2013.

\bibitem{MLYN16}
T.~L. Marzetta, E.~G. Larsson, H.~Yang, and H.~Q. Ngo, {\em Fundmentals of
  {M}assive {MIMO}}.
\newblock UK.: Cambridge Univ. Press, 2016.

\bibitem{Peaucelle-02-A}
D.~Peaucelle, D.~Henrion, and Y.~Labit, ``Users guide for {SeDuMi} interface
  1.03,'' 2002.

\bibitem{CT06}
T.~M. Cover and J.~A. Thomas, {\em Elements of Information Theory (second
  edition)}.
\newblock John Wileys \& Sons, 2006.

\bibitem{Adetal13}
A.~Adhikary, J.~Nam, J.~Ahn, and G.~Caire, ``Joint spatial division and
  multiplexing-the large-scale array regime,'' {\em IEEE Trans. Inf. Theory},
  vol.~59, pp.~6441--6463, Oct. 2013.

\bibitem{3GPP}
{3GPP}, ``{Study on channel model for frequencies from 0.5 to 100 GHz},'' 3rd
  Generation Partnership Project (3GPP), Sophia Antipolis Cedex, France, Tech.
  Rep. TR 38.901 (V14.2.0), Sep. 2017. [Online]. Available: http://
  www.3gpp.org/.

\bibitem{Jain84}
R.~Jain, D.-M. Chiu, and W.~R. Hawe, ``A quantitative measure of fairness and
  discrimination for resource allocation in shared computer systems,'' {\em
  Digital Equipment, Tech. Rep. DEC-TR-301}, Sept. 1984.

\end{thebibliography}

\end{document}